\documentclass[longbibliography,reprint,aps,prb,superscriptaddress,twocolumn,floatfix]{revtex4-2}
\usepackage[english]{babel}
\usepackage{tikz}
\usepackage{graphicx,graphics,grffile}%
\usepackage{dcolumn}%
\usepackage[abs]{overpic}
\usepackage[a4paper,top=2cm,bottom=2cm,left=2cm,right=2cm]{geometry}

\usepackage{amsmath,amssymb,amsfonts}
\usepackage{xcolor}
\usepackage[breaklinks=true,hidelinks]{hyperref}
\usepackage{braket}
\usepackage{bm}
\usepackage{listings}

\begin{document}

\title{A Majorana Formulation of Time-Dependent Two-Particle Reduced-Density-Matrix Dynamics
}
\author{Haoran Wang}
\email{haoran.wang-6@postgrad.manchester.ac.uk}
\affiliation{\noindent Department of Physics and Astronomy, University of Manchester, Manchester M13 9PL, UK}
\author{Alessandro Principi}
\email{alessandro.principi@manchester.ac.uk}
\affiliation{\noindent Department of Physics and Astronomy, University of Manchester, Manchester M13 9PL, UK}

\begin{abstract}
We introduce a time-dependent two-particle (TP) reduced-density-matrix algorithm for systems of interacting Majorana fermions. We close the BBGKY hierarchy at the two-particle level by reconstructing the three-particle reduced density matrix from lower-order correlations.
To stabilize the evolution, we impose a positivity projection on the two-particle density matrix while preserving chosen conserved quantities such as the energy. We benchmark the approach on quenches in the one-dimensional Hubbard model and on flux dynamics in the Kitaev honeycomb model. For the Hubbard quench, TP captures strong-coupling dynamics missed by Hartree--Fock (HF). For the Kitaev quenches, TP accurately reproduces the early-time flux dynamics in several regimes and provides a substantial improvement over HF when two-particle correlations are important. We further show that gauge coherence is essential for flux dynamics. The projection procedure however introduces an effective irreversibility, which affects the long-time dynamics.
\end{abstract}

\maketitle

\section{Introduction}
Solving a quantum many-body problem by directly manipulating the full wavefunction is a computationally hard problem. The dimension of the Hilbert space grows exponentially with particle number, making exact solutions impossible except for the smallest systems or for special integrable models. A longstanding strategy is therefore to replace the wavefunction by a reduced object that retains the information most relevant to the physical question being asked. Hartree--Fock (HF) theory is the simplest example of this idea: it compresses the many-body wavefunction into a Slater determinant, or equivalently a one-particle reduced density matrix. 

As efficient and useful as {HF theory} is, it discards correlations that can be important in strongly interacting problems. In a Mott insulator, energetic suppression of double occupancy causes insulating behavior. However, within {HF theory} the system can only acquire a gap by forming a symmetry-broken state, such as an antiferromagnet. This is because, under the {HF approximation}, the amplitude for double occupancy is decoupled into a product of single-particle occupation numbers: $\braket{\hat{n}_{i,\uparrow}\hat{n}_{i,\downarrow}}\rightarrow\braket{\hat{n}_{i,\uparrow}}\braket{\hat{n}_{i,\downarrow}}$. Hence, avoiding double occupancy forces either the spin-up or the spin-down occupation to vanish at each site $i$, naturally leading to an ordered state. This motivates a two-particle approximation as the next step: in addition to single-particle correlators, it retains all two-particle correlators, including the double-occupancy amplitudes.

The idea of using a two-particle description of many-electron systems was recognized long ago. For electronic Hamiltonians, which contain only single- and two-particle interactions, the ground-state energy is a linear functional of the two-particle reduced density matrix (2RDM). This observation motivated early attempts to determine the ground state directly by variational minimization over the 2RDM, without explicit reference to the many-electron wavefunction~\cite{mayer1955electron,tredgold1957density,coleman1963structure}. However, these approaches yielded energies below the exact ground-state energy, because not every antisymmetric two-particle density matrix corresponds to an electronic wavefunction. The admissible 2RDMs must satisfy the so-called N-representability conditions; otherwise the variational search includes unphysical density matrices and therefore gives spuriously low energies~\cite{coleman1963structure,garrod1964reduction,kummer1967nrepresentability}. Although the complete N-representability problem is highly nontrivial and, in general, computationally intractable, major progress was made by imposing certain necessary but insufficient conditions, which enabled practical variational 2RDM calculations through semidefinite programming~\cite{erdahl1978representability,nakata2001variational,mazziotti2002variational,zhao2004reduced,mazziotti2004realization,mazziotti2012twoelectron}.

The time-dependent reduced-density-matrix program is built on the Bogoliubov-Born-Green-Kirkwood-Yvon (BBGKY) hierarchy, originally formulated for classical statistical mechanics~\cite{yvon1935theorie,bogoliubov1946problems,kirkwood1946statistical,born1946general}. Closely related equation-of-motion hierarchies were later developed for quantum Green's functions~\cite{martin1959theory,zubarev1960double}. In the present work we solve the hierarchy for reduced-density-matrices, or equivalently equal-time Green's functions, and focus on closures at the two-particle level.

The two-particle (TP) reduced-density-matrix method was first applied in nuclear many-body problems, where the three-particle reduced density matrix is reconstructed from lower-order correlations in time-dependent density-matrix approaches~\cite{wang1985explicit,tohyama2020applications,schuck2016progress}. Later, it was applied to electronic problems~\cite{schaeferbung2008correlated,lackner2015propagating,lackner2017highharmonic,donsa2023nonequilibrium}. A central difficulty is that truncation can drive the reduced density matrices outside the representable set as the evolution proceeds, producing negative eigenvalues and numerical instabilities~\cite{akbari2012challenges}. So-called ``purification''~\cite{mazziotti2002purification} schemes were introduced to impose positivity on the reduced density matrices. However, purification breaks conservation laws, and therefore requires additional corrections~\cite{donsa2023nonequilibrium,pescoller2025projective}.

A main contribution of this work is a Majorana formulation of the time-dependent TP method, including its hierarchy closure and projective purification. This formulation allows us to treat fermion systems that do not conserve particle number as well as spin systems represented in terms of Majorana operators.
We formulate the hierarchy directly for fully antisymmetric one- and two-particle correlators, close it by reconstructing the three-particle reduced density matrix from lower-order correlations, and stabilize the resulting dynamics with a projective purification step that can be chosen to preserve selected conserved quantities. We benchmark the method against exact diagonalization and {HF} for a small one-dimensional Hubbard model quenched at time $t=0$ and for flux dynamics in a small Kitaev spin-liquid system. These examples show that retaining two-particle correlations substantially improves the description of strong-coupling transport and field-driven flux motion, while also making clear the main limitation of the present scheme: the repeated projective purification is intrinsically non-unitary and therefore introduces an effective irreversibility into the long-time dynamics.

\section{The algorithm}
We consider a fermionic system with $N$ available orbitals or lattice sites. We label the orbitals with $1, \ldots, N$ and assign them the annihilation operators $a_1,\ldots, a_{N}$. Therefore, the Hilbert space has dimension $2^N$. Since we will be considering spin as well as electronic systems, we will work in a framework that does not assume fermion number conservation. Therefore, we introduce $2N$ Majorana operators $m_1,\ldots, m_{2N}$ as follows:
\begin{align}
    m_{2n-1}&=a_n^\dagger + a_n \\
    m_{2n}&=-ia_n^\dagger+i a_n
    \label{eq:ferm2maj}
\end{align}
The Majorana operators are Hermitian and satisfy the anticommutation relation $\{m_i,m_j\}=2\delta_{ij}$. We consider a generic Hamiltonian which contains quadratic and quartic terms in Majorana basis.
We group all quadratic and quartic terms into $\hat{H}^{(2)}$ and $\hat{H}^{(4)}$, respectively, such that the Hamiltonian reads, schematically,
\begin{align}
    \hat{H}&=\hat{H}^{(2)}+\hat{H}^{(4)},
\end{align}
where
\begin{align}
    \hat{H}^{(2)}&=iH^{(2)}_{ij}m_im_j,
    \nonumber\\
    \hat{H}^{(4)}&=H^{(4)}_{ijkl}m_im_jm_km_l.
\end{align}
Here, summation over repeated indices is understood.
We use antisymmetric $H^{(2)}$ and $H^{(4)}$. This means diagonal terms such as $H^{(2)}_{ii}m_im_i=H^{(2)}_{ii}$ are not included. Note that such terms  produce only constant contributions to the Hamiltonian. Therefore, they affect neither the dynamics nor the ground-state energy and wave function.

To study the time evolution of observables, we solve the Heisenberg equations of motion for the {$k$-body antisymmetrized reduced density matrices (aRDMs). For mutually distinct indices, the first three are}
\begin{align} \label{eq:redDM_defs}
    M^{(1)}_{ij} &=\braket{m_i m_j}, \nonumber\\
    M^{(2)}_{ijkl} &=\braket{m_im_jm_km_l}, \nonumber\\
    M^{(3)}_{ijklpq} &=\braket{m_im_jm_km_lm_pm_q}.
    \nonumber\\
    \ldots
\end{align}
Here, the expectation value $\langle\ldots\rangle$ is taken over the instantaneous density matrix.
When any of the indices coincide, $M^{(k)}$ vanishes.

Working with aRDMs has a clear advantage. It avoids the need to enforce the so-called ``contraction-consistency constraints''~\cite{mazziotti2012twoelectron,pescoller2025projective,joost2022dsl}. 
These constraints need to be imposed when working with
non-antisymmetrized RDMs written in terms of  ordinary fermion creation and annihilation operators, $a_i^\dagger$ and $a_i$, respectively. For example, the RDMs $A_{ij}=\braket{a_i^\dagger a_j}$, $D_{ijkl}=\braket{a_i^\dagger a_j^\dagger a_k a_l}$ and $Q_{ijkl}=\braket{a_i^\dagger a_j a_k^\dagger a_l}$.
must satisfy the fermionic anticommutation identity $D_{ikjl}+Q_{ijkl}=A_{il}\delta_{jk}$. These equalities
must be enforced throughout the computation.
Avoiding this problem and working with aRDMs comes with a numerical cost. In systems with fermion number conservation, the memory size of the Majorana two-body aRDM $M^{(2)}$ is $(2N)^4$ (if it is stored as a dense array). 
In the conventional approach one can instead drop density matrices that break fermion number conservation, such as $\braket{a_i^\dagger a_j^\dagger a_k^\dagger a_l^\dagger}$, thereby reducing memory usage. The numerical advantage of non-antisymmetrized RDMs disappears when fermion number is not conserved. This is the case of the  Kitaev spin liquid problem we will consider below. Therefore, in our case aRDMs are the natural choice.

The Heisenberg equations of motion for the aRDMs
take the form of an infinite hierarchy. 
In fact, the equations of motion couple the $n$-body aRDM to the $n-1$- and $n+1$-body aRDMs via the quartic term in the Hamiltonian. The first three equations in the hierarchy
read
\begin{align}\nonumber
    \partial_t M^{(1)}_{ab}=&-8~\Upsilon_{ab}(H^{(2)}_{ai}M^{(1)}_{bi})
    \\
    \nonumber
    &+16i~\Upsilon_{ab}(H^{(4)}_{aijk}M^{(2)}_{bijk})
    \\
    \nonumber
    \partial_t M^{(2)}_{abcd}=&-16~\Upsilon_{abcd}(H^{(2)}_{ai}M^{(2)}_{bcdi}) \\
    \nonumber
    &-192i~\Upsilon_{abcd}(H^{(4)}_{abci}M^{(1)}_{di}) 
    \\
    \nonumber
    &+32i~\Upsilon_{abcd}(H^{(4)}_{aijk}M^{(3)}_{bcdijk})
	\\
    \nonumber
    \partial_t M^{(3)}_{abcdef}=&-24~\Upsilon_{abcdef}(H^{(2)}_{ai}M^{(3)}_{bcdefi}) \\
    \nonumber
    &+48i~\Upsilon_{abcdef}(H^{(4)}_{aijk}M^{(4)}_{bcdefijk}) 
    \\
    &-960i~\Upsilon_{abcdef}(H^{(4)}_{abci}M^{(2)}_{defi})
    \label{eqn:hierarchy}
\end{align}
where the antisymmetrizer $\Upsilon$ acts on a tensor $T$ as
\begin{equation}
    \Upsilon_{a_{1}\ldots a_n}(T_{a_1\ldots a_n})=\frac{1}{n!}\sum_P (-1)^P T_{a_{P_1}\ldots a_{P_n}}.
\end{equation}
Here, the sum is over all permutations of indices $a_1\ldots a_n$ of tensor $T$. In Eq.~(\ref{eqn:hierarchy}), summation is understood over repeated indices not appearing in $\Upsilon$.

Equation~(\ref{eqn:hierarchy}) is the equivalent of the usual BBGKY hierarchy for aRDMs written in Majorana representation~\cite{yvon1935theorie,bogoliubov1946problems,kirkwood1946statistical,born1946general,wang1985explicit,tohyama2020applications}.
In general, to solve the problem one needs to truncate the hierarchy at some level. If one truncates it at the first equation, decoupling $M^{(2)}_{ijkl}$ using Wick's theorem, the result is the HF approximation. Since the one-particle aRDM $M^{(1)}$ is very often insufficient to describe the physics of the problem, the HF approximation produces qualitatively wrong results. We will show that this is the case in the models studied in this manuscript. Therefore, to improve the approximation, we truncate at the level of the second equation by decoupling the {three-particle aRDM} $M^{(3)}$ in terms of the {one- and two-particle aRDMs} $M^{(1)}$ and $M^{(2)}$. The closure we employ corresponds to neglecting the connected three-particle cumulant, as in standard cumulant reconstruction schemes for higher-order aRDMs~\cite{colmenero1993general,nakatsuji1996density,mazziotti1999pursuit}. 

In passing, we note that the idea of decoupling higher-order quantities in terms of lower-order ones goes back to the early practices of terminating hierarchies of equations-of-motion for Green's functions, as in Refs.~\cite{bogolyubov1959retarded,zubarev1960double,tahirkheli1962use,callen1963green,hubbard1963electron,roth1969twopole}. Similar decoupling ideas were later applied to truncate the equation hierarchy for reduced density matrices in nuclear physics~\cite{wang1985explicit,tohyama2020applications,schuck2016progress} and quantum chemistry~\cite{nakatsuji1996density,yasuda1997densityii}.

We therefore decouple the three-particle aRDM $M^{(3)}$ as
\begin{align}
    \nonumber
    M^{(3)}_{abcdef}={}&15\Upsilon_{abcdef}
    \left(M^{(1)}_{ab}M^{(1)}_{cd}M^{(1)}_{ef}\right)
    \\
    \nonumber
    &+15\Upsilon_{abcdef}
    \left(M^{(2),\text{c}}_{abcd}M^{(1)}_{ef}\right)
    \\
    &+M^{(3),\text{c}}_{abcdef},
    \label{eq:Cdecouple}
\end{align}
where $M^{(3),\text{c}}$ is the connected part of the three-particle aRDM.

In Eq.~(\ref{eq:Cdecouple}), we also defined the connected part of $M^{(2)}$, $M^{(2),\text{c}}$ as
\begin{align}
    \nonumber
    &M^{(2),\text{c}}_{abcd}
    \\
    \nonumber
    =&M^{(2)}_{abcd}-(M^{(1)}_{ab}M^{(1)}_{cd}-M^{(1)}_{ac}M^{(1)}_{bd}+M^{(1)}_{ad}M^{(1)}_{bc})
    \\
    =&M^{(2)}_{abcd}-3\Upsilon_{abcd}(M^{(1)}_{ab}M^{(1)}_{cd})
\end{align}
Finally, $M^{(3),\text{c}}$ is the connected part of the three-particle aRDM $M^{(3)}$.

Our first approximation consists in dropping 
$M^{(3),\text{c}}$ entirely.
Using the antisymmetrizer $\Upsilon$, this amounts to rewriting Eq.~(\ref{eq:Cdecouple}) as
\begin{equation}
    M^{(3)}_{abcdef}\approx\Upsilon_{abcdef}(15M^{(2)}_{abcd}M^{(1)}_{ef}-30M^{(1)}_{ab}M^{(1)}_{cd}M^{(1)}_{ef})
    \label{eq:decouple}
\end{equation}
This defines the two-particle (TP) algorithm employed below.

A better approximation can be obtained by decoupling the three-particle connected part $M^{(3),\text{c}}$ in terms of $M^{(2)}$ and $M^{(1)}$, {\it i.e.}
\begin{align}\nonumber
    &M^{(3),\text{c}}_{abcdef}
    \\
    &\approx\frac{5}{4}\Upsilon_{abcdef}(M^{(2),\text{c}}_{abci}M^{(1)}_{ij}M^{(2),\text{c}}_{jdef}-3M^{(1)}_{ai}M^{(2),\text{c}}_{ibcj}M^{(2),\text{c}}_{jdef})
	\label{eq:pert}
\end{align}
The fermionic version of this formula as well as its derivation by perturbation theory can be found in~\cite{nakatsuji1996density,yasuda1997densityii,mazziotti1999pursuit}.
In this manuscript we will refer to the above equation as the ``perturbatively corrected decoupling''.
The corresponding perturbatively-corrected two-particle algorithm will be denoted with ``PTP'' below.

Direct integration of Eq.~(\ref{eqn:hierarchy}) with any three-particle aRDM closure condition [Eq.~(\ref{eq:decouple}) or Eq.~(\ref{eq:pert})] leaves the average energy unchanged. More generally, it leaves unchanged the expectation value of any conserved operator expressible as a sum of quadratic and quartic Majorana terms. To prove this, we schematically decompose one integration time step as two successive operations, which we call $\mathbf{R}$, and $\mathbf{E}$. They are superoperators that transform density operators into density operators, which we define as follows. 

Given any density operator $\rho(t)$, the (nonlinear) superoperator $\mathbf{R}$ (which stands for reconstruct) turns it into a state $\mathbf{R}[\rho(t)]$ that has vanishing three- and higher-order connected aRDMs, in the following way. First, $M^{(1)}(t)$ and $M^{(2)}(t)$ for $\rho(t)$ are constructed according to Eq.~(\ref{eq:redDM_defs}). Then, all higher-order aRDMs can be found according to, e.g., Eq.~(\ref{eq:decouple}) and its generalizations to higher-order ones. One can then ``reconstruct'' the density matrix schematically as
\begin{equation}
    \mathbf{R}[\rho(t)] = \frac{1}{2^N}\left[1+\frac{1}{2!}\sum_{ab}M^{(1)}_{ab}(t)m_b m_a %
    +\ldots\right],
    \label{eq:schem_recons}
\end{equation}
which satisfies $\text{Tr}\big[\mathbf{R}[\rho(t)]\big]=1$.
The one- and two-particle aRDMs obtained from $\mathbf{R}[\rho(t)]$ are, by construction, $M^{(1)}(t)$ and $M^{(2)}(t)$, respectively. {Higher-order aRDMs} are given by the decoupling rule applied. In particular, since $M^{(1)}(t)$ and $M^{(2)}(t)$ are the exact aRDMs for the state $\rho(t)$, the reconstructed state $\mathbf{R}[{\rho}(t)]$ has the same averages as the original state for any operator expressible as a sum of products of two or four Majorana operators. The Hamiltonian, being a quartic operator, falls within this category. 

The superoperator $\mathbf{E}$, which stands for evolve, performs the exact time evolution for one time step $dt$, i.e., $\mathbf{E}[\rho(t)]=U(dt)\rho(t)U^\dagger(dt)$, where $U$ is the unitary time-evolution operator. The combined operation of $\mathbf{E}$ and $\mathbf{R}$ yields a density operator $\mathbf{E}[\mathbf{R}[\rho(t)]]$. The one- and two-particle aRDMs for $\mathbf{E}[\mathbf{R}[\rho(t)]]$ can be found exactly by integrating the first two equations of the hierarchy Eq.~(\ref{eqn:hierarchy}), using the same closure condition chosen for $\mathbf{R}$. This holds because, by definition of $\mathbf{R}$, the three-particle aRDM for the state $\mathbf{R}[\rho(t)]$ is exactly given by the decoupling formula used. Therefore, the hierarchy Eq.~(\ref{eqn:hierarchy}) gives the exact derivatives for $M^{(1)},M^{(2)}$. Since energy stays the same after the operation $\mathbf{R}$, and $\mathbf{E}$ performs exact time evolution, their combined operation does not alter the expectation value of energy.

Integrating Eqs.~(\ref{eqn:hierarchy}) with the decoupling Eq.~(\ref{eq:decouple}) leads however to divergent results. This is because, given the one- and two-particle correlations $M^{(1)}$ and $M^{(2)}$, the reconstructed state $\tilde{\rho}$ in general has negative eigenvalues, violating the principle that probabilities must be positive. In principle, one could diagonalize $\tilde{\rho}$ and remove its negative eigenvalues. However, this is clearly unfeasible for large systems. Instead, a number of positivity constraints can be imposed on the aRDMs. These are known as $N$-representability conditions~\cite{coleman1963structure,mazziotti2012twoelectron,mazziotti2002purification,pescoller2025projective}. We first note that
\begin{equation}
    \braket{Q^\dagger Q}\geq 0
\end{equation}
holds for an arbitrary operator $Q$; in particular, upon taking $Q=\lambda_{ij}m_im_j$, the above condition is equivalent to the positive semidefiniteness of the matrix
\begin{equation}
    F_{ij,kl}=\braket{(m_i m_j)^\dagger m_k m_l},
\end{equation}
viewed as a matrix whose rows are indexed by the pair $(i,j)$ and whose columns are indexed by the pair $(k,l)$.

To prevent a divergence of the solutions, we project $F$ onto the space of positive semidefinite matrices. This process is sometimes called ``purification''~\cite{mazziotti2002purification,pescoller2025projective}. We diagonalize $F$ by
\begin{equation}
    F_{\alpha\beta}=\sum_k U_{\alpha k}^\dagger f_k U_{k\beta} 
\end{equation}
where $\alpha=(i,j)$ and $\beta=(k,l)$ are the flattened indices and $f_k$ are the eigenvalues of $F$. The negative part of $F$ is then defined as
\begin{equation}
    F^\text{neg}_{\alpha\beta}=\sum_{k,f_k<0} U_{\alpha k}^\dagger f_k U_{k\beta} 
\end{equation}
One could be tempted to simply subtract $F^\text{neg}$ from $F$.
Unfortunately, this would break conservation laws. This is avoided as follows.
We consider any conserved operator $O$ that can be written as sums of one- and two-body terms, {\it i.e.} $\hat{O}=\sum_{ij} O^{(2)}_{ij}m_i m_j + \sum_{ijkl} O^{(4)}_{ijkl}m_i m_j m_k m_l$. The average $\braket{O}$ can be written in terms of $F$ as
\begin{equation}
    \braket{O}=\sum_{ij}O^{(2)}_{ij}F_{1i;1j}+\sum_{ijkl}O^{(4)}_{ijkl}F_{ji;kl},
\end{equation}
where the index 1 can be replaced by any other index ($F$ satisfies $F_{ui;uj}=\braket{m_i m_u m_u m_j}=\braket{m_i m_j}$ for any index $u$). We then replace $F$ with ${\tilde F} \equiv F - {\tilde F}^\text{neg}$, where ${\tilde F}^\text{neg}$ is obtained from $F^\text{neg}$ by setting to zero all entries ${\tilde F}^\text{neg}_{ui;uj}$ and ${\tilde F}^\text{neg}_{ji;kl}$ for which $O^{(2)}_{ij}$ and $O^{(4)}_{ijkl}$ are nonzero, respectively.
In this way, the expectation value
\begin{equation}
\sum_{ij}O^{(2)}_{ij}{\tilde F}_{ui;uj}+\sum_{ijkl}O^{(4)}_{ijkl}{\tilde F}_{ji;kl}
\end{equation}
coincides with 
$\braket{O}$. Repeating this for any conserved operator $O$ ensures that the average value $\braket{O}$ stays conserved during the time evolution.

The downside of this procedure is that the negative eigenvalues of $F$ are not completely removed. In principle, the above procedure ($F\rightarrow F-\tilde{F}^\text{neg}$) can be iterated indefinitely to ensure complete removal of negative eigenvalues of $F$.
However, we find that complete positivity is unnecessary to ensure stability of the calculation. Furthermore, we find that performing too many such projections can lead to early thermalization, owing to the non-unitary nature of projections. In subsequent sections, we will describe in detail the origin of these errors.

OpenAI Codex (GPT-5.6 Sol and GPT-6 Astra) assisted with developing and checking parts of the numerical analysis, and with preparing plotting code. The code for TP, HF and exact diagonalization was almost entirely human-written; ChatGPT was used to rearrange that code to perform calculations based on human instructions, as well as writing independent verification code.

\section{Hubbard molecule}
\label{sect:Hubbard}

As a simple test we apply the two-particle algorithm to a four-site Hubbard molecule. We first introduce the Hubbard model in terms of fermion operators. We then transform into the Majorana basis using Eqs.~(\ref{eq:ferm2maj}) and solve the two-particle hierarchy~(\ref{eqn:hierarchy}) with the closure Eq.~(\ref{eq:decouple}) (TP algorithm) or the perturbatively corrected Eq.~(\ref{eq:pert}) (PTP algorithm). We will compare the results for the two closure conditions and benchmark them against the exact and HF results.

The Hamiltonian for a one-dimensional four-site Hubbard chain reads
\begin{equation}
    H_{\text{hub}}=J_h\sum_{i,\alpha} (a_{i,\alpha}^\dagger a_{i+1,\alpha} + a_{i+1,\alpha}^\dagger a_{i,\alpha}) + U\sum_i \hat{n}_{i,\uparrow} \hat{n}_{i,\downarrow}
    \label{eq:Hubbard_model}
\end{equation}
where $a_{i,\alpha}$ is the electron annihilation operator. 
The four lattice sites are labeled by $i$, 
while $\alpha=\uparrow,\downarrow$ is the spin label. In Eq.~(\ref{eq:Hubbard_model}), we defined $\hat{n}_{i,\alpha}=a_{i,\alpha}^\dagger a_{i,\alpha}$ as the number operator on site $i$ and for a given spin projection $\alpha$. For notational convenience, in this section we set $J_h$ as the unit of energy and $\hbar=1$. Hence, time is in units of $J_h^{-1}$.

In the repulsive Hubbard model, Coulomb interactions cause the electrons to avoid each other. Such correlations are at least partially captured at the level of the two-particle aRDM, {\it i.e.} by the truncations discussed in Eqs.~(\ref{eq:decouple}) and~(\ref{eq:pert}). By contrast, a truncation to the single-particle aRDM, which corresponds to the HF approximation, fails to describe the behaviour of electrons, in particular in the limit of large $U$. This can be understood on general grounds as follows.

In the limit of large $U$, correlators of form $\braket{\hat{n}_{i,\uparrow}\hat{n}_{i,\downarrow}}$ are expected to be close to zero in the ground state, since particles repel each other. Within the HF approximation, such correlators are decoupled as $\braket{\hat{n}_{i,\uparrow}\hat{n}_{i,\downarrow}} = \braket{\hat{n}_{i,\uparrow}}\braket{\hat{n}_{i,\downarrow}}$. The requirement of $\braket{\hat{n}_{i,\uparrow}\hat{n}_{i,\downarrow}} \approx 0$ implies that either $\braket{\hat{n}_{i,\uparrow}}$ or $\braket{\hat{n}_{i,\downarrow}}$, or both, must be close to zero. This naturally leads to symmetry broken states.
However, if the ground state of the system does not break any symmetry, then the HF approximation fails.

To see this, consider the singlet state of two electrons on different sites $i=1, 2$, {\it i.e.} $(\ket{\uparrow \downarrow}-\ket{\downarrow \uparrow})/\sqrt{2}$. Since each site hosts one electron, this state has zero onsite Coulomb energy. However, the expectation values $\braket{a^\dagger_{i,\alpha}a_{i,\alpha}}=0.5$ for both $i=1,2$ and $\alpha=\uparrow,\downarrow$, therefore the HF part of its onsite Coulomb energy is $U/4$. This example shows that higher-order correlations play a vital role in the description of correlated states, and therefore a truncation to the one-particle aRDM is insufficient.

\begin{figure}
    \centering
    \includegraphics*[width=0.9\linewidth]{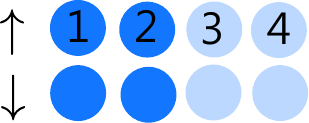}
    \caption{Setup of the four-site one-dimensional Hubbard model studied in Sec.~\ref{sect:Hubbard}. Four electrons, two spin-up and two spin-down, are initially placed on sites 1 and 2 (dark blue). Sites 3 and 4 (light blue) are initially empty.}
	\label{fig:hubbard_setup}
\end{figure}

Our setup includes four sites and four electrons, as shown in Fig.~\ref{fig:hubbard_setup}. At time $t=0$ we place all four electrons on the two leftmost sites 1 and 2. We monitor the occupation $n_{1,\uparrow}(t)=\braket{\hat{n}_{1,\uparrow}}$. We use a time step $dt = 0.01~J_h^{-1}$.

\begin{figure}
    \centering
    \begin{overpic}[width=1.\linewidth]{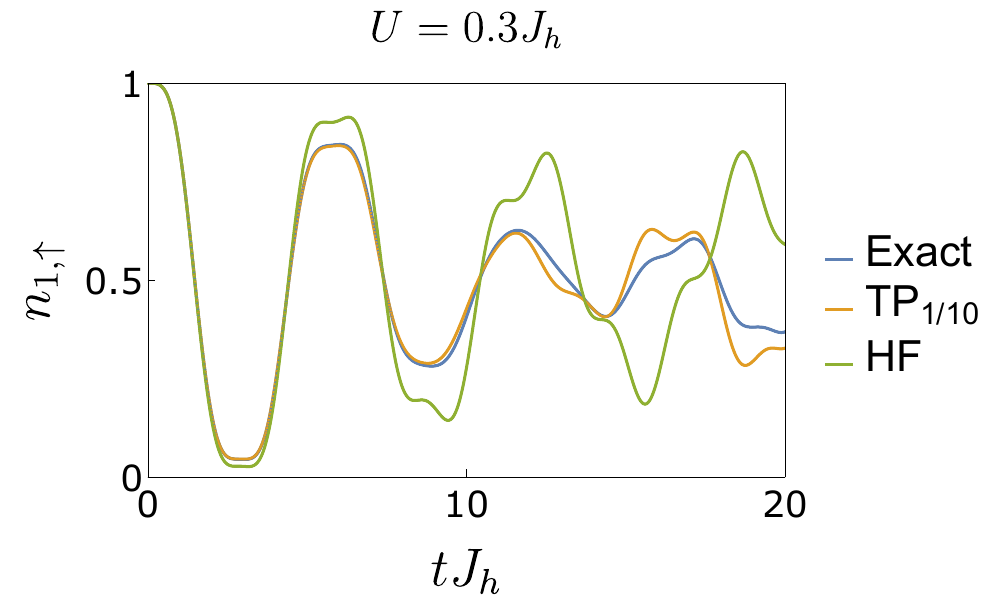}
        \put(0,0){(a)}
    \end{overpic}
    \vspace{0.2cm} \\
    \begin{overpic}[width=1.\linewidth]{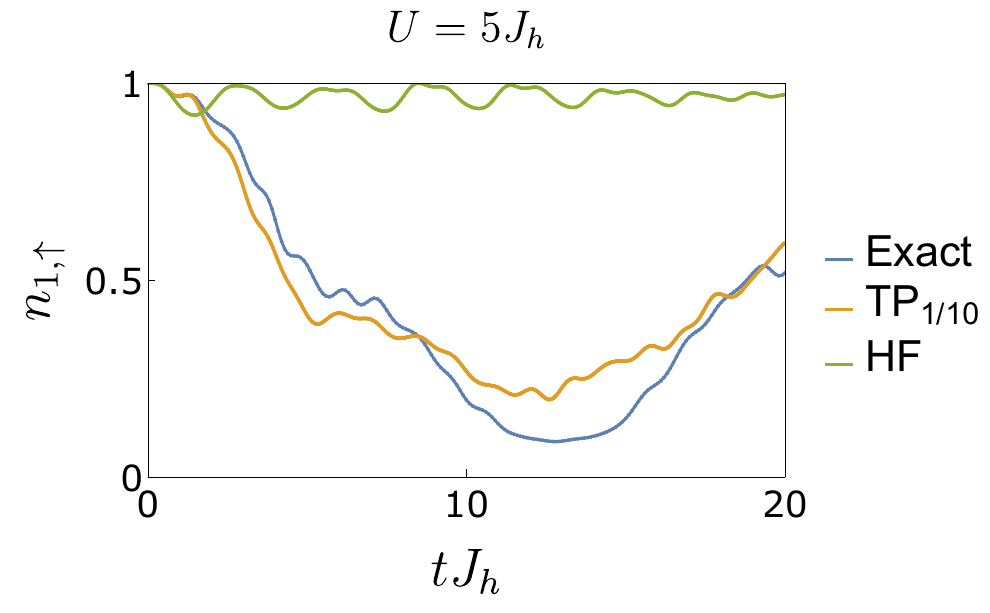}
        \put(0,0){(b)}
    \end{overpic}
    \caption{Time evolution of the expectation value of the occupation number $\braket{\hat{n}_{1,\uparrow}}$ for the four-site Hubbard model following a quantum quench. Three algorithms are compared: exact diagonalization, TP$_{1/10}$, and HF. Here TP$_{1/10}$ denotes that one projection is performed every ten time steps.
    Panel (a) shows the weak-interaction case $U=0.3J_h$.
    Panel (b) shows the strong-interaction case $U=5J_h$.
    }
    \label{fig2}
\end{figure}

In Fig.~\ref{fig2} we compare results for weak ($U=0.3J_h$) and strong ($U=5J_h$) interactions for three algorithms: exact diagonalization, HF, and TP. In this figure, TP is implemented by performing one projection every ten time steps. 
At the low interaction strength, $U=0.3J_h$ [panel (a)], both TP and HF accurately reproduce the dynamics at short times $(t<5J_h^{-1})$. Over longer times HF starts to oscillate with amplitudes larger than those of the exact results, whereas the amplitudes for TP remain small and track the exact result. At $U=5J_h$ [panel (b)], TP remains in good agreement with the exact results for the whole duration of the simulation. On the other hand, the HF result remains always very close to $\langle {\hat n}_{1,\uparrow}\rangle\approx 1$.

The fact that electrons become almost immobile under HF can be understood as follows. Since $U\gg J_h$, the energy is dominated by the potential energy part $U\sum_i \braket{\hat{n}_{i,\uparrow} \hat{n}_{i,\downarrow}}$, which under the HF approximation is replaced by $U\sum_i \braket{\hat{n}_{i,\uparrow}}\braket{\hat{n}_{i,\downarrow}}$. Let us represent the initial configuration of the system graphically as
\begin{equation}
\ket{\psi(0)}=\begin{tikzpicture}[baseline=-0.0ex, x=0.8em, y=0.8em]
    \draw[rounded corners=1.5pt, line width=0.35pt]
        (-0.45,-0.4) rectangle (3.45,1.4);

    \fill (0,1) circle (0.17);
    \fill (1,1) circle (0.17);
    \draw[line width=0.35pt] (2,1) circle (0.17);
    \draw[line width=0.35pt] (3,1) circle (0.17);

    \fill (0,0) circle (0.17);
    \fill (1,0) circle (0.17);
    \draw[line width=0.35pt] (2,0) circle (0.17);
    \draw[line width=0.35pt] (3,0) circle (0.17);
\end{tikzpicture}
~.
\label{eq:config1}
\end{equation}
Here, the upper row represents the occupation of spin-up electrons, while the lower one shows the occupation of spin-down ones. Filled black (empty white) dots represent filled (empty) sites.
The configuration shown in Eq.~(\ref{eq:config1}) has energy $2U$. Now suppose the electrons are trying to reach the final state
\begin{equation}
\ket{\psi(t_f)}=\begin{tikzpicture}[baseline=-0.0ex, x=0.8em, y=0.8em]
    \draw[rounded corners=1.5pt, line width=0.35pt]
        (-0.45,-0.4) rectangle (3.45,1.4);

    \fill (0,1) circle (0.17);
    \fill (2,1) circle (0.17);
    \draw[line width=0.35pt] (1,1) circle (0.17);
    \draw[line width=0.35pt] (3,1) circle (0.17);

    \fill (0,0) circle (0.17);
    \fill (2,0) circle (0.17);
    \draw[line width=0.35pt] (1,0) circle (0.17);
    \draw[line width=0.35pt] (3,0) circle (0.17);
\end{tikzpicture}
~.
\end{equation}
During the time evolution, there must be an intermediate time $t^*$ when the occupation numbers are
\begin{equation}\ket{\psi({t^*})}\sim
\begin{tikzpicture}[baseline=1.2ex, x=2em, y=2em]
    \draw[rounded corners=1.5pt, line width=0.35pt]
        (-0.45,-0.4) rectangle (3.45,1.4);

    \node at (0,1) {$1$};
    \node at (1,1) {$\frac{1}{2}$};
    \node at (2,1) {$\frac{1}{2}$};
    \node at (3,1) {$0$};

    \node at (0,0) {$1$};
    \node at (1,0) {$\frac{1}{2}$};
    \node at (2,0) {$\frac{1}{2}$};
    \node at (3,0) {$0$};
\end{tikzpicture}
~.
\end{equation}
The HF approximation gives an energy $U+U/4+U/4=1.5U$ for this state. However, the exact wavefunction is given by a superposition
\begin{equation}
\ket{\psi(t^*)}\approx\frac{1}{\sqrt{2}}\left(\begin{tikzpicture}[baseline=-0.0ex, x=0.8em, y=0.8em]
    \draw[rounded corners=1.5pt, line width=0.35pt]
        (-0.45,-0.4) rectangle (3.45,1.4);

    \fill (0,1) circle (0.17);
    \fill (1,1) circle (0.17);
    \draw[line width=0.35pt] (2,1) circle (0.17);
    \draw[line width=0.35pt] (3,1) circle (0.17);

    \fill (0,0) circle (0.17);
    \fill (1,0) circle (0.17);
    \draw[line width=0.35pt] (2,0) circle (0.17);
    \draw[line width=0.35pt] (3,0) circle (0.17);
\end{tikzpicture}+\begin{tikzpicture}[baseline=-0.0ex, x=0.8em, y=0.8em]
    \draw[rounded corners=1.5pt, line width=0.35pt]
        (-0.45,-0.4) rectangle (3.45,1.4);

    \fill (0,1) circle (0.17);
    \fill (2,1) circle (0.17);
    \draw[line width=0.35pt] (1,1) circle (0.17);
    \draw[line width=0.35pt] (3,1) circle (0.17);

    \fill (0,0) circle (0.17);
    \fill (2,0) circle (0.17);
    \draw[line width=0.35pt] (1,0) circle (0.17);
    \draw[line width=0.35pt] (3,0) circle (0.17);
\end{tikzpicture}\right)
.
\end{equation}
Clearly, this superposition has energy $2U$. Thus, HF neglects the correlated part of the potential energy, $E_c=U/2$. However, since energy is conserved in the HF time evolution, all intermediate states must have energy $2U$.
The only possibility, for HF, is to stop electrons from moving, otherwise conservation of energy would be violated.

TP, on the other hand, keeps track of the full electron potential energy, because energy is a two body operator. The correlation energy $E_c$ is included directly in the evolved density matrices. In particular, for the intermediate state $\ket{\psi(t^*)}$ one has that $\braket{\hat{n}_{2,\uparrow}\hat{n}_{2,\downarrow}}=\braket{\hat{n}_{3,\uparrow}\hat{n}_{3,\downarrow}}=1/2$; while HF replaces these correlators with products of single-particle ones, TP includes them directly. Therefore, TP gives the correct potential energy $U+U/2+U/2=2U$.

\begin{figure}
	\centering
	\begin{tabular}{@{}c@{}}
		\begin{overpic}[width=1.\linewidth]
			{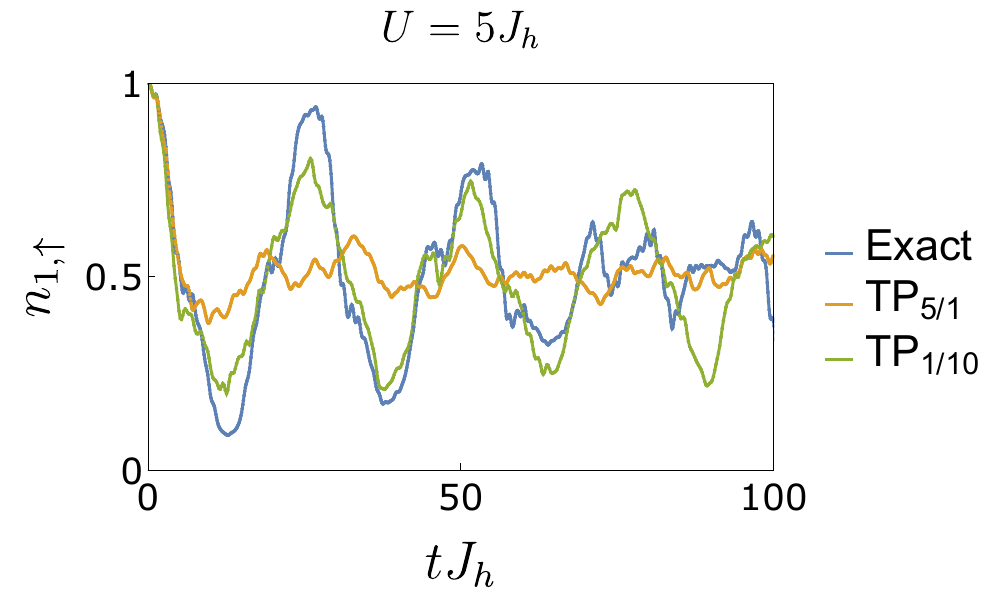}
		\end{overpic} 
	\end{tabular}
	\caption{
    TP evolution for different projection rates and compared with the exact result. Here TP$_{5/1}$ denotes five projections every time step, while TP$_{1/10}$ denotes one projection every ten time steps.  TP$_{5/1}$ converges quickly to the average value, whereas TP$_{1/10}$ tracks the exact result for significantly longer times.}
	\label{fig3}
\end{figure}

In Fig.~\ref{fig3}, we analyze the impact of the projection on the time evolution. We compare performing five projections for every time step (denoting the corresponding TP algorithm with TP$_{5/1}$) with one projection every ten time steps. In the latter case, the algorithm is denoted as TP$_{1/10}$. We also plot the exact result, which oscillates with a period $\sim 25~{J_h}^{-1}$ around the equilibrium value $\langle {\hat n}_{1,\uparrow}\rangle=1/2$ with a slowly decaying amplitude. TP$_{1/10}$ agrees well with the exact result. In contrast, the time evolution under TP$_{5/1}$ converges to the equilibrium value in a very short time scale $\approx 10~J_h^{-1}$, and then oscillates around it with a very small amplitude. We therefore conclude that performing too many projections per unit time can lead to early thermalization, which is a consequence of the inherent non-unitarity of the projection process. 

\begin{figure}
    \centering
	\begin{tabular}{@{}c@{}}
		\begin{overpic}[width=1.\linewidth]
			{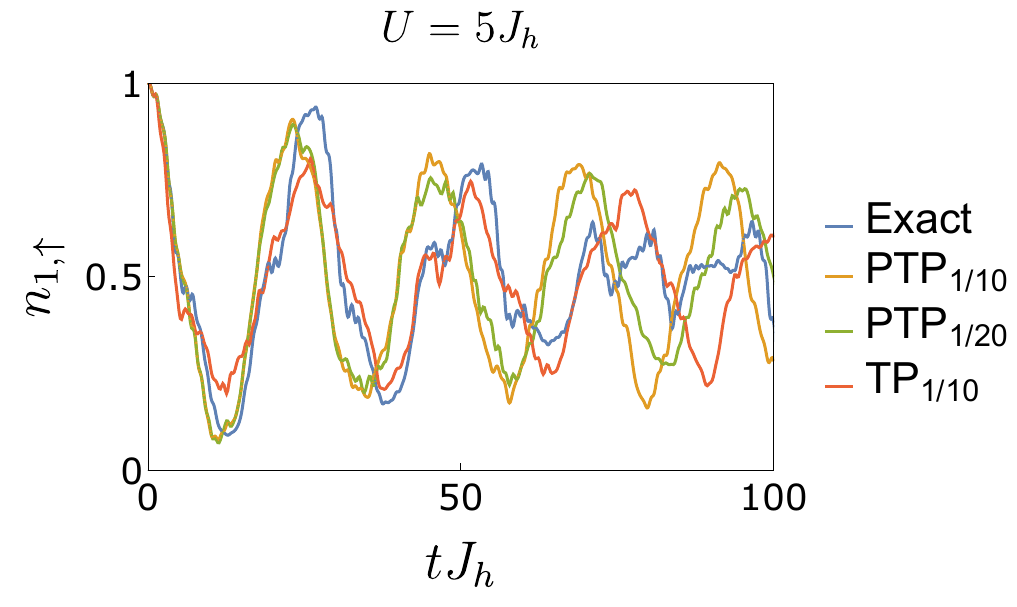}
		\end{overpic} 
	\end{tabular}
	\caption{
		Impact of the perturbative correction on TP. PTP$_{1/n}$ uses one projection every $n=10, 20$ time steps. TP$_{1/10}$, as before, denotes TP with one projection every 10 time steps. Perturbative corrections improve the agreement with the exact result at early times, but produce too fast oscillations at later times.}
        \label{fig4}
\end{figure}

In Fig.~\ref{fig4}, we evaluate the performance of perturbatively corrected TP (PTP) where we use the closure~(\ref{eq:pert}). 
In this figure, the algorithm is implemented with one projection every ten time steps (PTP$_{1/10}$) and one projection every twenty time steps (PTP$_{1/20}$). Both PTP algorithms agree better than TP$_{1/10}$ with the exact result at early times, $t\lesssim 20J_h^{-1}$. However, the subsequent oscillations appear to have a higher frequency than those of both the exact and TP$_{1/10}$ algorithms.
We therefore conclude that, while perturbative corrections improve the agreement of the approximate short-time dynamics with the exact result, they worsen the later time evolution compared to TP algorithms.

\section{Kitaev model}
\label{sect:Kitaev}
\begin{figure}
    \includegraphics[width=0.7\linewidth]{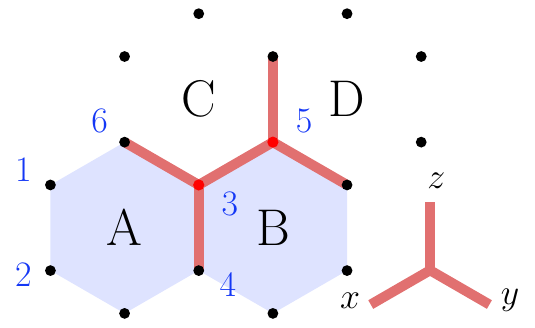}
    \caption{Schematic of the four-plaquette Kitaev model studied in Sec.~\ref{sect:Kitaev}. The plaquettes are labelled with the letters $A,B,C,D$. We initialize the system by placing fluxes at plaquettes $A$ and $B$. In this manuscript, we denote this configuration as $AB$.
    \label{fig:kitaev_system} The schematic was prepared with OpenAI Codex assistance in writing the TikZ drawing code.}
\end{figure}

Next we consider application of TP to the isotropic ferromagnetic Kitaev honeycomb model, which hosts an exactly solvable spin liquid. We consider a four-plaquette (16-site) honeycomb lattice with open boundary conditions, as shown in Fig.~\ref{fig:kitaev_system}. We label the four plaquettes as $p=A,B,C,D$ as illustrated there. The Hamiltonian of the system is ${\hat H}={\hat H}_{\rm K}+{\hat H}_h$, where
\begin{align} \label{eq:Kitaev_Hamiltonian}
    {\hat H}_K = - J\sum_{\langle j,k\rangle} \sigma_j^{\alpha_{jk}}\sigma_k^{\alpha_{jk}} 
\end{align}
is the unperturbed Kitaev Hamiltonian~\cite{Kitaev_2006}, which couples the $\alpha_{jk}$ spin component along the nearest-neighbour bond $\langle j,k\rangle$. Here, $\sigma_i^\alpha$ denotes the $\alpha$-th Pauli matrix acting on the spin-$1/2$ Hilbert space at site $i$, while $\alpha_{jk}$ takes the values $x,y,z$ depending on the direction of the bond $\langle j,k\rangle$. With reference to Fig.~\ref{fig:kitaev_system}, $\alpha_{jk}$ is such that $\alpha_{3,6}=y$, $\alpha_{3,5}=x$ and $\alpha_{3,4}=z$. Bonds parallel to these have the same Ising-like spin couplings. The local Zeeman coupling to the external field $\bm{h}_j$ (which can depend on the position of site $j$) is
\begin{align}
{\hat H}_h = \sum_{j} \bm{h}_j\cdot\bm{\sigma}_j,
\end{align}
Hereafter, we scale energies with the Kitaev coupling $J$ and we set $\hbar=1$. Therefore, times are in units of $J^{-1}$. 

The model~(\ref{eq:Kitaev_Hamiltonian}) can be exactly solved by mapping Pauli matrices into Majorana operators. Following Ref.~\cite{Kitaev_2006}, we introduce four Majorana operators $b^x_j$, $b^y_j$, $b^z_j$ and $c_j$ at each lattice site $j$, such that $\sigma^\alpha_j=ib_j^\alpha c_j$. In the Majorana representation, the Kitaev Hamiltonian becomes
\begin{equation} \label{eq:Kitaev_Hamiltonian_Majorana}
    {\hat H}_{K} = i J \sum_{\langle j,k\rangle} u^{\alpha_{jk}}_{jk} c_j c_k.
\end{equation}
where all $u^{\alpha_{jk}}_{jk}=i b^{\alpha_{jk}}_jb^{\alpha_{jk}}_k$ commute with ${\hat H}_{K}$ and are therefore constants of motion of the unperturbed Kitaev model. The sum counts each nearest-neighbour bond once, with $j$ on the even sublattice and $k$ on the odd sublattice. Their eigenvalues are $\pm 1$. Therefore, the Hamiltonian~(\ref{eq:Kitaev_Hamiltonian_Majorana}) describes the dynamics of Majorana fermions $c$, hereafter called matter fermions, in the presence of a $\mathbb{Z}_2$ gauge potential~\cite{Kitaev_2006}.

The Majorana representation doubles the Hilbert space of the original Kitaev model. The component of any state belonging to the unphysical part of the Hilbert space is removed by acting on the wave function with an appropriate projector operator~\cite{Kitaev_2006}. To define this, we introduce the gauge transformation operator $D_j= b^x_jb^y_jb^z_j c_j$, which satisfies $D_j=1$ for physical states (for which it coincides with $-i \sigma_j^x \sigma_j^y  \sigma_j^z$) and $D_j=-1$ for unphysical ones~\cite{Kitaev_2006}. The sought projector is then constructed as 
\begin{equation}
    P=\prod_j \frac{1+D_j}{2}
    \label{eq:projector}
\end{equation}
where the product is over all lattice sites $j$.

The Kitaev Hamiltonian in Eq.~(\ref{eq:Kitaev_Hamiltonian}) possesses a set of conserved quantities $W_p = \pm 1$, conventionally named ``fluxes'' or ``visons''. Here, $p$ is one of the hexagonal plaquettes. In the Majorana representation, fluxes are given by
\begin{equation}
    W_p=\prod_{\langle j,k\rangle}u^{\alpha_{jk}}_{jk},
\end{equation}
where $\langle j,k\rangle$ runs over the six boundary bonds of plaquette $p$, each oriented from the even to the odd sublattice, with $u_{kj}^{\alpha_{jk}}=-u_{jk}^{\alpha_{jk}}$. Owing to Lieb's theorem~\cite{lieb1994flux}, the ground state of the Kitaev model must satisfy $W_p=1$ for any plaquette $p$. This state is said to be ``flux free''.
For calculations in the enlarged Majorana space, we choose a representative of the flux-free sector by setting $u_{ij}^{\alpha_{ij}}=-1$ when $i$ and $j$ are even and odd, respectively. Other gauge-related link assignments represent the same physical flux sector.
When $W_p=-1$, the plaquette $p$ is said to be threaded by a flux.

We distinguish a physical flux sector from a gauge sector. A gauge sector is specified by a simultaneous assignment of link eigenvalues $\{u_{jk}^{\alpha_{jk}}\}$ and therefore depends on the gauge choice. Acting with $D_j$ reverses the three link variables incident on site $j$ but leaves every flux $W_p$ unchanged. A flux sector is instead specified only by the gauge-invariant set of plaquette eigenvalues $\{W_p\}$ and contains all gauge sectors related by the local transformations $D_j$. Thus, with reference to Fig.~\ref{fig:kitaev_system} we use labels such as $AB$ for the physical flux sector containing fluxes only in plaquettes $A$ and $B$, not to denote particular assignments of link variables. We will need flux projectors $\Pi_s$, which projects onto the flux sector $s$, defined in the following way. The flux sector $AB$ has $W_A=W_B=-1$ and $W_C=W_D=1$, therefore
\begin{equation}
    \Pi_{AB}=\frac{1-W_A}{2}\frac{1-W_B}{2}\frac{1+W_C}{2}\frac{1+W_D}{2},
    \label{eq:flux_proj}
\end{equation}
and likewise for the other flux sectors. 

In the following numerical calculations, we use a time step $dt=0.01J^{-1}$. TP is implemented with one projection every ten time steps. We consider the motion of fluxes driven by a magnetic field. We initialize the system in an eigenstate with two flux excitations threading plaquettes $A$ and $B$, as shown in Fig.~\ref{fig:kitaev_system}. Within the chosen gauge, a representative of this $AB$ flux sector is obtained from the flux-free representative by flipping $u_{34}^{\alpha_{34}}$.%
We first analyze the effects of a magnetic field term applied to a single site ($j=3$). We consider two distinct field directions, {\it i.e.}, along $y$ or $z$. Therefore, we study the effects of $\hat{H}_h=h\sigma_3^y$ and $\hat{H}_h=h\sigma_3^z$. For simplicity, we call these the $3y$ and $3z$ magnetic fields, respectively, where ``3'' denotes the site at which the field is applied. With reference to Fig.~\ref{fig:kitaev_system}, the $3y$ field causes the flux at plaquette $A$ to hop onto plaquette $C$, and therefore the system oscillates between the flux configurations $AB$ and $BC$. The $3z$ field causes hopping between the $AB$ and flux-free ($FF$) configurations.

\begin{figure}
	\centering
	\begin{tabular}{@{}c@{}}
		\begin{overpic}[width=0.9\linewidth]
			{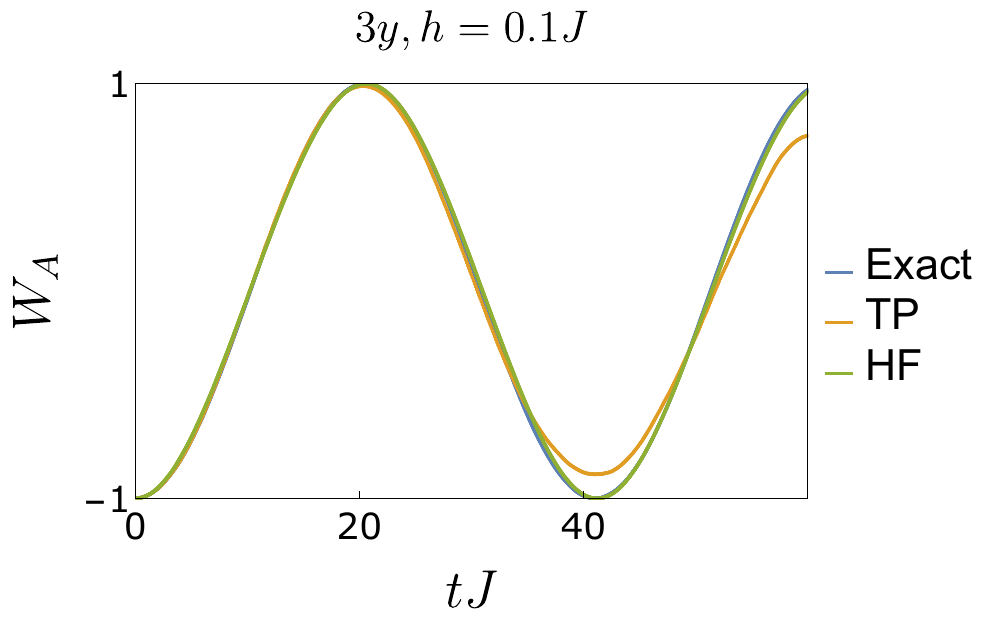}
			\put(0,0){(a)}
		\end{overpic}
        \vspace{0.2cm} \\
		\begin{overpic}[width=0.9\linewidth]
			{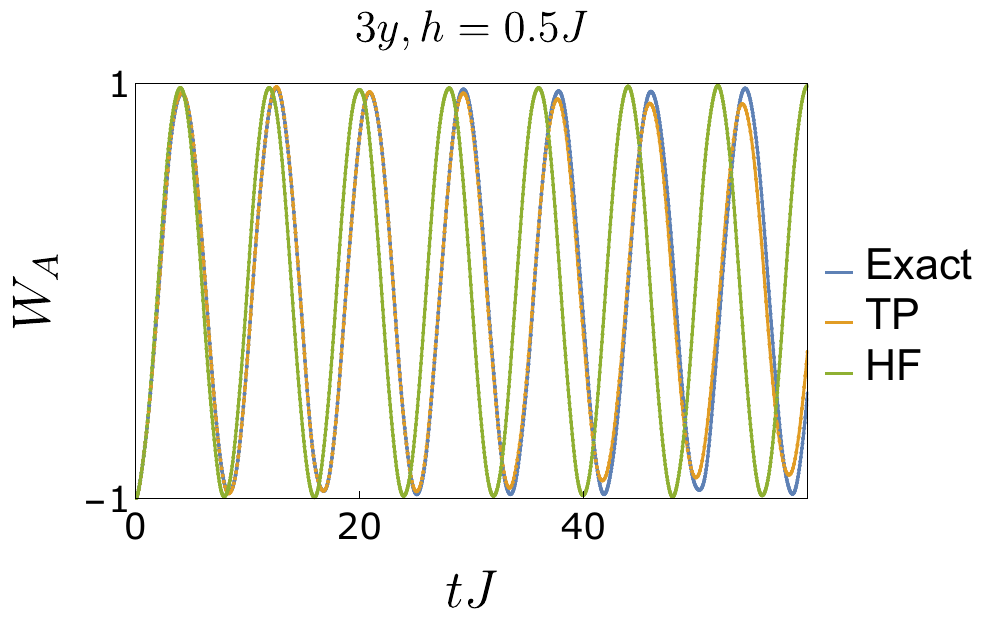}
			\put(0,0){(b)}
		\end{overpic}
	\end{tabular}
	\caption{
		The time evolution of a four-plaquette Kitaev model following the sudden switch on of a magnetic field in the $y$ direction applied to site $3$
        ($3y$ quantum quench).
        See Fig.~\ref{fig:kitaev_system} for a definition of the lattice. In this case, the system oscillates between the $AB$ and $BC$ flux sectors.
        The dynamics obtained with the HF and TP algorithms is compared with the exact one. Here, TP is implemented with one projection every ten time steps.
        Panel (a) Time evolution of $W_A$ for a magnetic field $h=0.1J$. TP gives near-exact agreement at times $t\lesssim40$, but shows signs of relaxation. The HF flux remains close to the exact result. 
        Panel (b) At high field $h=0.5J$, both TP and HF predict qualitatively correct flux dynamics. TP predicts an accurate oscillation frequency but suffers from early relaxation, while HF has a slightly higher frequency than the exact result.}
	\label{fig:kitaev1}
\end{figure}

\begin{figure}
	\centering
	\begin{tabular}{@{}c@{}}
		\begin{overpic}[width=0.9\linewidth]
			{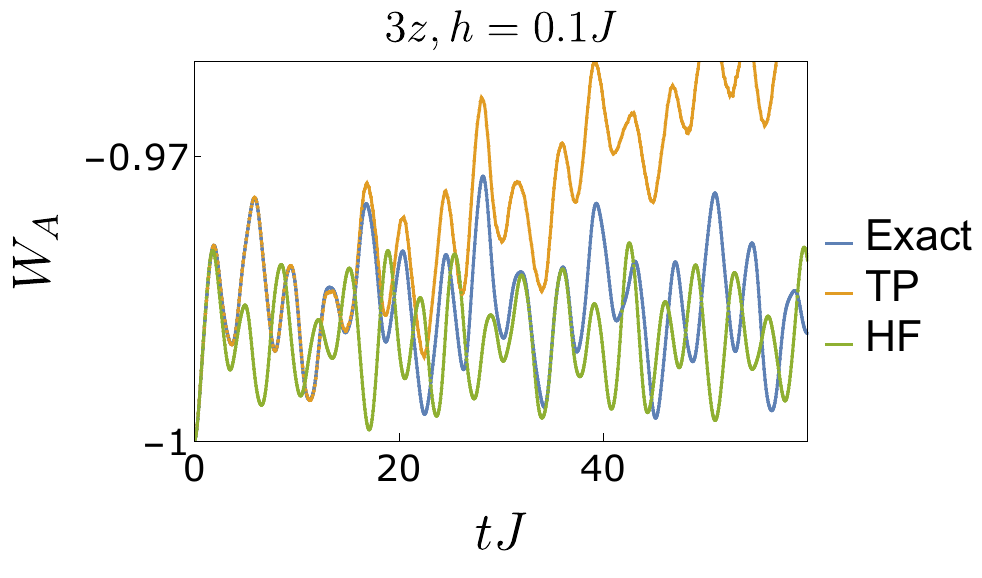}
			\put(0,0){(a)}
		\end{overpic}
        \vspace{0.2cm} \\
		\begin{overpic}[width=0.9\linewidth]
			{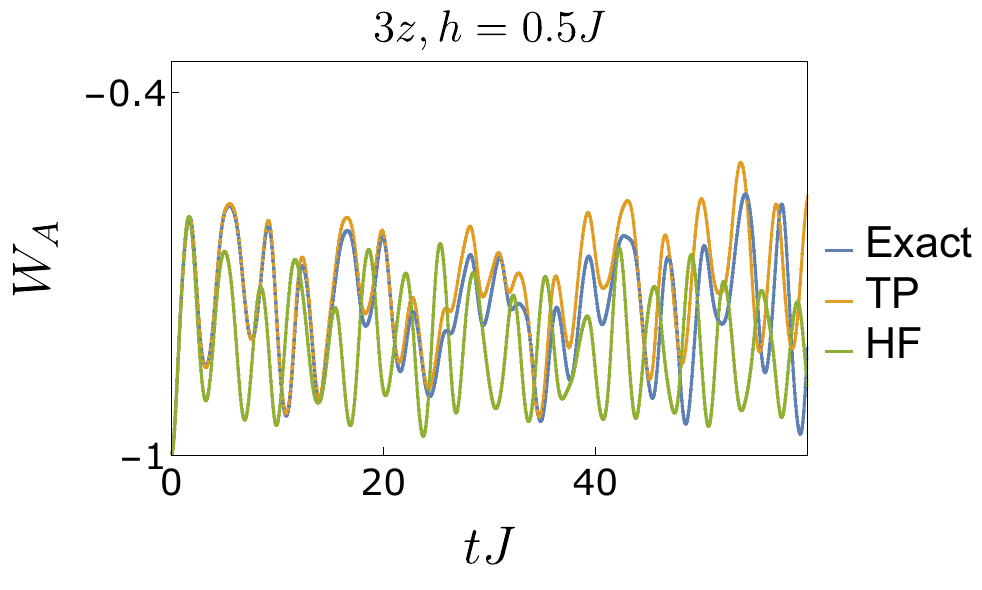}
			\put(0,0){(b)}
		\end{overpic}
	\end{tabular}
	\caption{
		The time evolution of a four-plaquette Kitaev model following the sudden switch on of a magnetic field in the $z$ direction applied to site $3$
        ($3z$ quantum quench).
        See Fig.~\ref{fig:kitaev_system} for a definition of the lattice. In this case, the system oscillates between the $AB$ and $FF$ (flux free) flux sectors.
        The dynamics obtained with the HF and TP algorithms is compared with the exact one. Here, TP is implemented with one projection every ten time steps.
        Panel (a) For low field $h=0.1J$, the oscillation amplitudes are significantly smaller than for the $3y$ quench [Fig.~\ref{fig:kitaev1}(a)]. TP gives near-exact agreement with the exact result at early times.
        Panel (b) At high field $h=0.5J$, TP again gives near-exact agreement at early times.}
	\label{fig:kitaev3z}
\end{figure}

Figures~\ref{fig:kitaev1}(a) and~(b) show the time evolution of the expectation value of the flux operator through plaquette A, $\braket{W_A}$, following the sudden switch-on of the $3y$ field. There, the performances of HF and TP algorithms are compared with exact diagonalisation. All three methods reproduce the first low-field peak, but the TP oscillation amplitude decreases after $t\sim30J^{-1}$, likely because of purification. At high field, TP continues to follow the exact dynamics reasonably well with a progressively reduced amplitude, whereas HF remains nearly undamped but slightly overestimates the oscillation frequency.

We note that the exact-diagonalization result has an especially simple behavior: the time evolution of the average flux number resembles the Rabi oscillation of a two-state system. By contrast, the evolution of $\braket{W_A}$ when a $3z$ field is suddenly switched on is much more complicated, see Figs.~\ref{fig:kitaev3z}(a) and~(b).
The dynamics presents smaller and more irregular oscillations. TP tracks their timing accurately at early times but develops an upward drift after $t\gtrsim20J^{-1}$, which becomes more pronounced at high field. Conversely, HF departs earlier, with peaks and valleys shifted toward earlier times, again indicating an overestimate of the characteristic frequencies.

The behavior observed for the $3y$ field can be captured by the following simple model. The $3y$ Zeeman field moves the flux from plaquette $A$ to plaquette $C$, so that the initial $AB$ configuration is coupled to the $BC$ configuration. We define the corresponding projected Kitaev Hamiltonians as $H_{AB}=\Pi_{AB}H_K\Pi_{AB}$ and $H_{BC}=\Pi_{BC}H_K\Pi_{BC}$, where $\Pi_s$ is a flux sector projector as defined in Eq.~(\ref{eq:flux_proj}).  The state of the system during the time evolution driven by $\hat{H}_h=h\sigma_3^y$ can be written as 
\begin{equation}
    \ket{\psi(t)}=\ket{\psi_{AB}(t)}+\ket{\psi_{BC}(t)}
    \label{eq:Kitaev_split_state}
\end{equation}
where $\ket{\psi_{AB}(t)}=\Pi_{AB}\ket{\psi(t)}$ and $\ket{\psi_{BC}(t)}=\Pi_{BC}\ket{\psi(t)}$. The Schr\"odinger equations for these two states are
\begin{align}
    i\partial_t \ket{\psi_{AB}(t)}&= H_{AB} \ket{\psi_{AB}(t)} + h\sigma_3^y \ket{\psi_{BC}(t)}, \notag\\
    i\partial_t \ket{\psi_{BC}(t)}&= H_{BC} \ket{\psi_{BC}(t)} + h\sigma_3^y \ket{\psi_{AB}(t)}
    \label{eq:twostate1}
\end{align}
Exact diagonalization results show that, throughout the time evolution, $\ket{\psi_{AB}(t)}$ and $\ket{\psi_{BC}(t)}$ remain close to the ground states of $H_{AB}$ and $H_{BC}$, which we denote by $\ket{AB}$ and $\ket{BC}$, respectively. %
For simplicity, we choose the phases of $\ket{AB}$ and $\ket{BC}$ such that the overlap $\braket{AB|\sigma_3^y|BC}=\braket{BC|\sigma_3^y|AB}=v$ is a real number.
Therefore, we write
\begin{align}
    \ket{\psi_{AB}(t)}&=\alpha_{AB}(t)\ket{AB}+\ket{\eta_{AB}(t)},%
    \notag\\
    \ket{\psi_{BC}(t)}&=\alpha_{BC}(t)\ket{BC}+\ket{\eta_{BC}(t)}%
\end{align}
where $\ket{\eta_{AB}}$ and $\ket{\eta_{BC}}$ are orthogonal to $\ket{AB}$ and $\ket{BC}$, respectively.
The remainders $\ket{\eta_{AB}(t)}$ and $\ket{\eta_{BC}(t)}$ lie in the $AB$ and $BC$ flux sectors, respectively, and are orthogonal to the ground states $\ket{AB}$ and $\ket{BC}$.
More generally, for any flux sector $s$ with weight $p_s(t)=\langle\psi(t)|\Pi_s|\psi(t)\rangle>0$, we define the normalized sector-projected state and its sector-ground-state fidelity as
\begin{equation}
|\widetilde\psi_s(t)\rangle=
\frac{\Pi_s|\psi(t)\rangle}{\sqrt{p_s(t)}},
\qquad
F_s(t)=|\langle s|\widetilde\psi_s(t)\rangle|^2 .
\label{eq:effective_model_validity}
\end{equation}
Here $\ket{s}$ is the dynamically-reachable ground state in flux sector $s$. We find $\ket{s}$ by mapping the initial state into flux sector $s$ with the appropriate magnetic-field operators and then evolving the resulting state in imaginary time under $H_s=\Pi_sH_K\Pi_s$ until convergence.
The quantity $1-F_s(t)$ is the matter-excitation weight within that sector. An effective model retaining one ground state per sector is valid only when $F_s(t)\simeq1$ in every appreciably occupied retained sector and the total probability outside the retained flux sectors is negligible. %

Using $H_{AB}\ket{AB}=E_{AB}\ket{AB}$ and $H_{BC}\ket{BC}=E_{BC}\ket{BC}$, %
Eqs.~(\ref{eq:twostate1}) give the equations
\begin{align}
    i\dot{\alpha}_{AB}&=E_{AB}\alpha_{AB}+hv\alpha_{BC}
    +h\braket{AB|\sigma_3^y|\eta_{BC}}, \notag\\
    i\dot{\alpha}_{BC}&=E_{BC}\alpha_{BC}+hv\alpha_{AB}
    +h\braket{BC|\sigma_3^y|\eta_{AB}}
\end{align}
The effective two-state approximation is obtained by dropping $\ket{\eta_{AB}}$ and $\ket{\eta_{BC}}$, which is precisely the approximation $F_{AB},F_{BC}\approx1$. This gives
\begin{align}
    i\dot{\alpha}_{AB}&\approx E_{AB}\alpha_{AB}+hv\alpha_{BC} \notag \\
    i\dot{\alpha}_{BC}&\approx E_{BC}\alpha_{BC}+hv\alpha_{AB}
    \label{eq:twostate2}
\end{align}
The energies $E_{AB}$ and $E_{BC}$ are almost the same: $E_{AB}\approx-10.9157J$ and $E_{BC}\approx-10.9108J$. Replacing them by $E\approx E_{AB}\approx E_{BC}$, and applying the initial condition $\alpha_{AB}(0)=1,\alpha_{BC}(0)=0$, we have the solution
\begin{align}
    \alpha_{AB}(t)&=e^{-iEt}\cos(hvt) \notag \\
    \alpha_{BC}(t)&=-ie^{-iEt}\sin(hvt)
\end{align}
and the flux $W_A$ on plaquette A is given by
\begin{equation}
    \braket{W_A(t)}=|\alpha_{BC}(t)|^2-|\alpha_{AB}(t)|^2=-\cos(2hvt)
\end{equation}
Exact results show $v=\braket{BC|\sigma_3^y|AB}\approx 0.77$, which at the magnetic field $h=0.1J$ gives an angular frequency $2hv\approx0.154 J$. This is in good agreement with Fig.~\ref{fig:kitaev1}(a). In fact, fitting the exact result with $-\cos(\omega t)$ results in an angular frequency $\omega\approx 0.152 J$. At $h=0.5J$, we have $2hv\approx0.77 J$, which also closely agrees with Fig.~\ref{fig:kitaev1}(b), where an angular frequency $\omega\approx 0.75 J$ is found.

For the $3z$ field, the system oscillates between the $AB$ and $FF$ flux sectors. Applying the same two-state approximation gives
\begin{equation}
    \braket{W_A(t)}=-\frac{\Delta^2+4h^2u^2\cos(\sqrt{\Delta^2+4h^2u^2}t)}{\Delta^2+4h^2u^2}
\end{equation}
Here $u=|\langle FF|\sigma_3^z|AB\rangle|\approx0.195$ and $\Delta=E_{FF}-E_{AB}\approx0.553J$. At $h=0.1J$, the two-state expression predicts $\langle W_A\rangle_{\max}\approx-0.990$, which has the correct small scale but does not reproduce the irregular time dependence observed in Fig.~\ref{fig:kitaev3z}(a). In fact, up to an arbitrary phase,
\begin{equation}
\begin{aligned}
\sigma_3^z|AB\rangle
&\approx0.195|FF\rangle\\
&\quad+0.981|\eta_{FF}\rangle,
\end{aligned}
\end{equation}
where $\langle FF|\eta_{FF}\rangle=0$.
This in turn implies that only about $3.8\%$ of the state $\sigma_3^z|AB\rangle$ lies parallel to the ground state of the $FF$ sector. Thus, the two-state model fails because matter-excited components dominate the projected state.

\begin{figure}
    \centering
    \begin{tabular}{@{}c@{}}
        \begin{overpic}[width=0.86\linewidth]
            {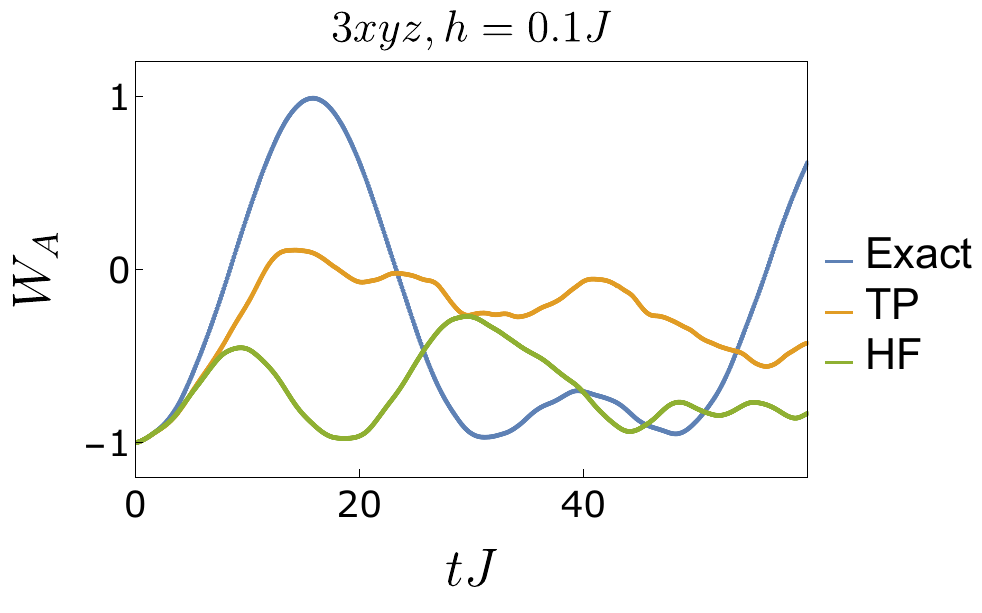}
            \put(0,0){(a)}
        \end{overpic}
        \vspace{0.2cm} \\
        \begin{overpic}[width=0.86\linewidth]
            {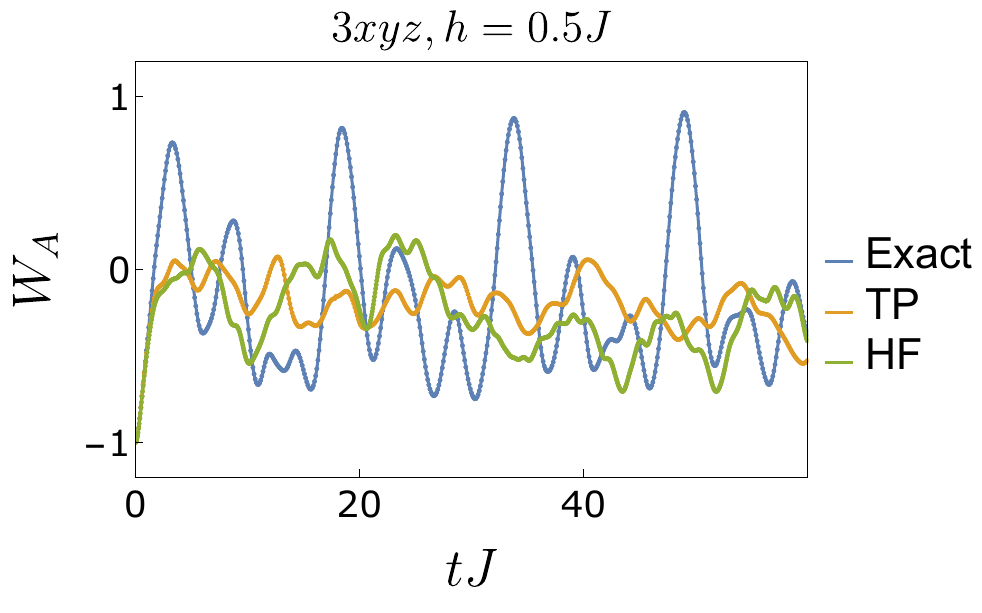}
            \put(0,0){(b)}
        \end{overpic}
        \vspace{0.2cm} \\
        \begin{overpic}[width=0.95\linewidth]
            {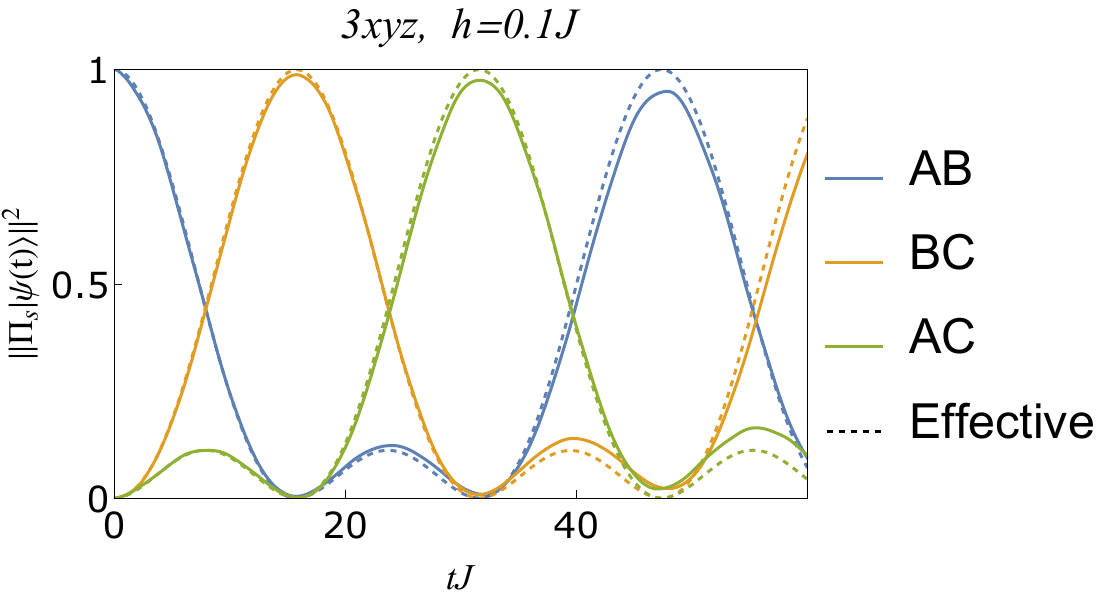}
            \put(0,0){(c)}
        \end{overpic}
    \end{tabular}
    \caption{
    The time evolution of a four-plaquette Kitaev model following the sudden switch on of a magnetic field with components in all three directions applied to site $3$ ($3xyz$ quantum quench).
    See Fig.~\ref{fig:kitaev_system} for a definition of the lattice. 
    {Panels (a) and (b) compare the exact, TP, and HF dynamics for $h=0.1J$ and $h=0.5J$, respectively.}
    Here and throughout the Kitaev section, TP is implemented with one representability projection every ten time steps.
    Panel (c) compares the exact weights of the three dominant flux sectors (solid lines) with the three-state effective model of Eq.~(\ref{eq:three_state_Kitaev}) (dashed lines) for $h=0.1J$. The 
    $FF$ contribution is not shown.}
    \label{fig:kitaev3xyz_results}
\end{figure}

\begin{figure}
    \centering
    \begin{tabular}{@{}c@{}}
        \begin{overpic}[width=0.86\linewidth]
            {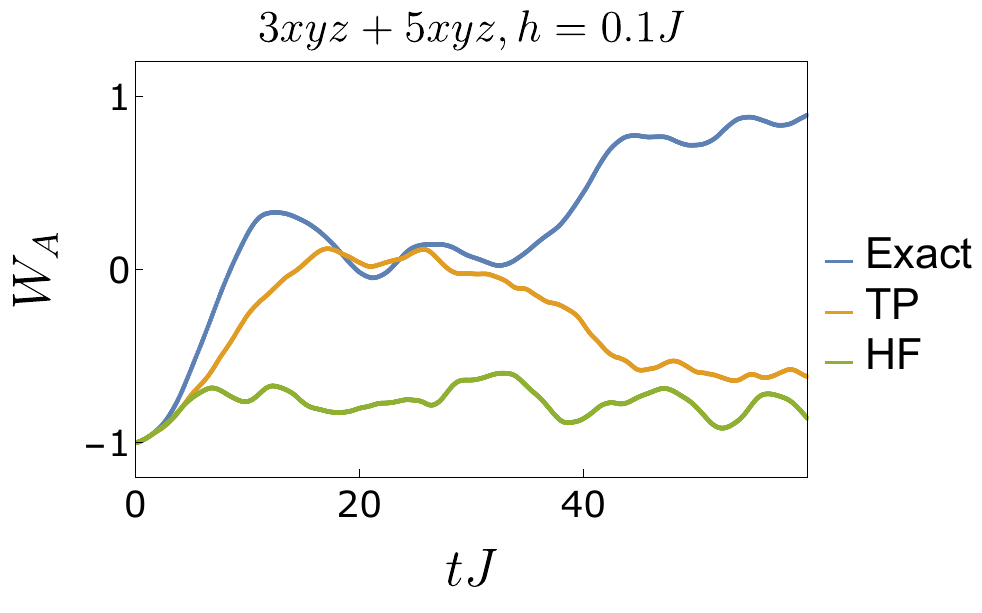}
            \put(0,0){(a)}
        \end{overpic}
        \vspace{0.2cm} \\
        \begin{overpic}[width=0.86\linewidth]
            {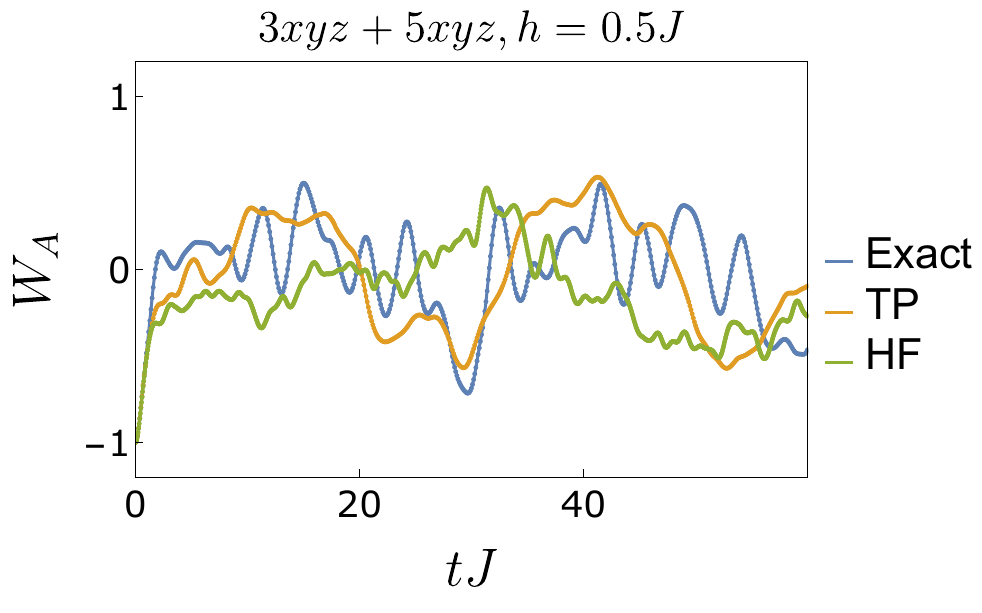}
            \put(0,0){(b)}
        \end{overpic}
        \vspace{0.2cm} \\
        \begin{overpic}[width=0.95\linewidth]
            {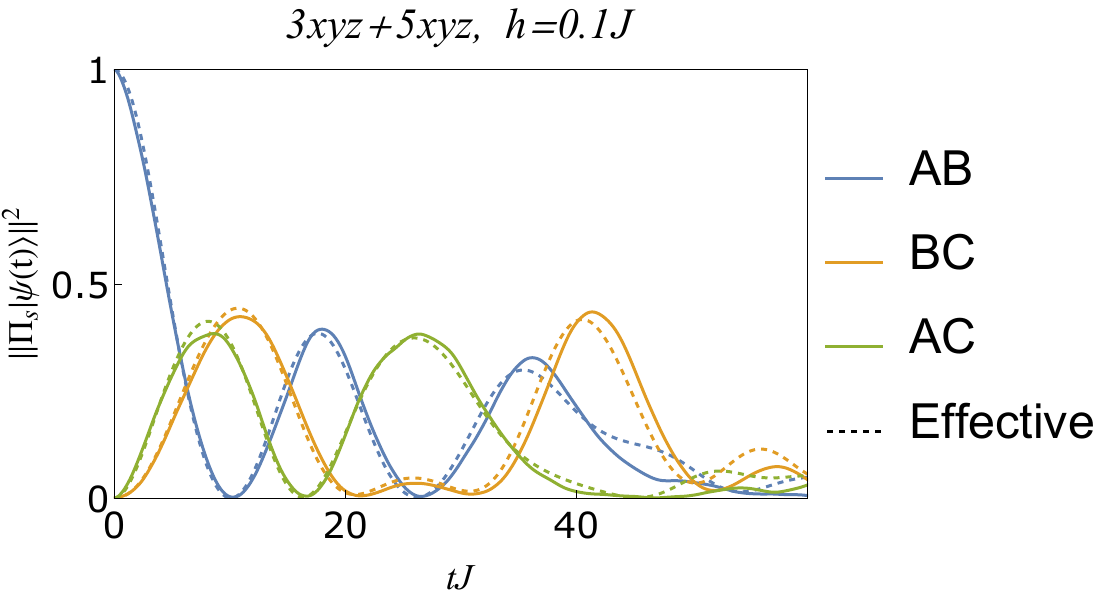}
            \put(0,0){(c)}
        \end{overpic}
    \end{tabular}
    \caption{The time evolution of a four-plaquette Kitaev model following the sudden switch on of a magnetic field with components in all three directions applied to both sites $3$ and $5$ ($3xyz+5xyz$ quantum quench).
    See Fig.~\ref{fig:kitaev_system} for a definition of the lattice. 
    {Panels (a) and (b) compare the exact, TP, and HF dynamics for $h=0.1J$ and $h=0.5J$, respectively.}
    TP uses one representability projection every ten time steps.
    Panel (c) compares selected exact two-flux-sector weights (solid lines) with the six-state effective model of Eq.~(\ref{eq:six_state_Kitaev}) (dashed lines) for $h=0.1J$. The $FF$ and $ABCD$ sectors are omitted from this low-field model because their combined exact weight is negligible over the interval shown. }
    \label{fig:kitaev3xyz5xyz_results}
\end{figure}

In Figs.~\ref{fig:kitaev3xyz_results} and~\ref{fig:kitaev3xyz5xyz_results}, we present the dynamics following a sudden quench of a magnetic field with components in all three directions. When applied to site 3, the field is $H_h=h\sum_{\alpha}\sigma_3^\alpha$, which we call the $3xyz$ field. Conversely, when applied to both sites 3 and 5, it is $H_h=h\sum_{\alpha}(\sigma_3^\alpha+\sigma_5^\alpha)$, and we call the $3xyz+5xyz$ field.
In both equations $\alpha$ takes the values $x,y$ and $z$. As before, we initialize the system in the ground state of the flux sector $AB$. We turn on the magnetic field $H_h$ at time $t=0$. For the $3xyz$ field, the system reaches four flux sectors $FF,AB,AC,BC$. For the $3xyz+5xyz$ field, the system reaches all eight possible flux sectors: the flux free sector $FF$, all two-flux sectors $AB,AC,\ldots,CD$, and the sector $ABCD$ where all four plaquettes are flipped.

For the $3xyz$ field, at $h=0.1J$ the exact result shows that the flux oscillates between $W_A=-1$ and $W_A=1$. 
Over the duration $0<t<60J^{-1}$ it reaches one maximum, and is on track to reach the second maximum just after $t>60J^{-1}$. It also reaches a much smaller peak in between the two maxima, at $t\sim40J^{-1}$.%
Figure~\ref{fig:kitaev3xyz_results}(a) shows that TP captures the initial direction of flux transfer and rises substantially above HF, but both approximations underestimate the first maximum and miss the later revival. %
At $h=0.5J$ [panel (b)], HF and TP reproduce the rapid initial departure from $W_A=-1$ and the broad scale of the subsequent fluctuations, but neither follows the 
exact pattern of fluctuations.

The low-field exact dynamics in Fig.~\ref{fig:kitaev3xyz_results}(c) is dominated by the three flux sectors $AB$, $BC$, and $AC$. The sector-ground-state fidelity $F_s$ is high for all three retained sectors, suggesting an effective three-state description. On the contrary, the weight of $FF$ remains at the percent level and therefore this state is omitted from the effective model below. The exact wavefunction is well approximated by
\begin{equation}
    |\psi(t)\rangle\approx \alpha_{AB}(t)|AB\rangle+\alpha_{BC}(t)|BC\rangle+\alpha_{AC}(t)|AC\rangle .
\end{equation}
Here $|AB\rangle$, $|BC\rangle$, and $|AC\rangle$ are the ground states of the respective flux sectors. Their phases are chosen such that the amplitudes are
\begin{align} \label{eq:j_123_choices}
j_1=\braket{AB|H_h|BC}&=0.07651J,\\
j_2=\braket{AB|H_h|AC}&=0.07662J,\\
ij_3=\braket{BC|H_h|AC}&=0.07651iJ,
\end{align}
{\it i.e.} $j_1$, $j_2$ and $j_3$ are real and positive.
The 
factor of $i$ in the last of~(\ref{eq:j_123_choices}) indicates an effective Berry phase of $\pi/2$ 
when hopping in the loop $AB\rightarrow BC \rightarrow AC \rightarrow AB$. 
We then solve the effective Hamiltonian evolution
\begin{equation} \label{eq:three_state_Kitaev}
    i\partial_t\begin{pmatrix}
        \alpha_{AB}(t) \\\alpha_{BC}(t) \\\alpha_{AC}(t)
    \end{pmatrix}=
    \begin{pmatrix}
        0 & j_1 & j_2 \\
        j_1 & 0 & ij_3 \\
        j_2 & -ij_3 & 0
    \end{pmatrix}\begin{pmatrix}
        \alpha_{AB}(t) \\\alpha_{BC}(t) \\\alpha_{AC}(t)
    \end{pmatrix}
\end{equation}
where we have ignored the energy differences between $\ket{AB},\ket{BC}$ and $\ket{AC}$ because they proved to be small. The solution to the above equation is shown by the dashed lines in Fig.~\ref{fig:kitaev3xyz_results}(c), which closely track the exact results. Both curves share an important feature: the system cycles through three sectors $AB$, $BC$ and $AC$, and while it is in transit between two of them, it only has a small component in the third sector. 

The three-state effective model does not reproduce the $3xyz$ dynamics at the larger field $h=0.5J$. Its limitation is that it retains only the ground state of each flux sector, whereas the exact state, projected into a given sector, contains a substantial matter-excited component and therefore is not well approximated by that sector's ground state. The high-field comparison is shown in Appendix~\ref{app:effective_high_field}.

In Fig.~\ref{fig:kitaev3xyz5xyz_results} we present the $3xyz+5xyz$ quench results. At $h=0.1J$, the exact flux $W_A$ first rises above zero near $t\sim10J^{-1}$ and increases again after $t\sim40J^{-1}$. Although the field connects all eight flux sectors, the combined exact weight of the $FF$ and $ABCD$ sectors is negligible on the scale of the displayed low-field dynamics. We therefore retain the six two-flux sectors and define
\begin{equation}
\label{eq:six_state_Kitaev}
(H_{\mathrm{eff}})_{ss'}=E_s\delta_{ss'}+\braket{s|H_h|s'}.
\end{equation}
Here $s,s'\in\{AB,AC,AD,BC,BD,CD\}$ and $|s\rangle$ is the ground state of flux sector $s$. We initialize the effective model in $|AB\rangle$. Figure~\ref{fig:kitaev3xyz5xyz_results}(c) shows selected $AB$, $BC$, and $AC$ weights and demonstrates that this six-state model captures the broad low-field dynamics accurately.

At $h=0.5J$, by contrast, the effective model fails because the projected state contains substantial matter-excited components within the flux sectors; see Appendix~\ref{app:effective_high_field}.

For the low-field $3xyz+5xyz$ quench, Fig.~\ref{fig:kitaev3xyz5xyz_results}(a), TP follows the initial rise of $W_A$ much more closely than HF, but turns downward after the first oscillations while the exact flux grows again at late times.  At high field [panel (b)], both approximations produce fluctuations of the correct broad scale, although neither retains the phases and recurrence pattern of the exact curve.

\subsection{Gauge invariance}
The comparisons above show that retaining independent two-particle correlations improves substantially on HF, but does not make TP reliable for the $3xyz$ and $3xyz+5xyz$ quenches.  We argue that the remaining difficulty is enforcing gauge invariance.  The physical spin state is gauge invariant, whereas the enlarged Majorana Hilbert space contains several gauge-related representatives of the same flux configuration.  Physical transition amplitudes are coherent sums over these representatives.

The $3xyz$ quench gives a concrete example.  Let $\bm u_3=(u_{36},u_{35},u_{34})$ denote the three gauge links incident on site 3 and choose $\bm u_3=(+1,+1,-1)$ as the initial $AB$ representative.  The direct transition $AB\rightarrow BC$ generated by $\sigma_3^y$ ends in $(-,+,-)$, whereas the indirect path $AB\rightarrow AC\rightarrow BC$, generated by $\sigma_3^x$ followed by $\sigma_3^z$, ends in $(+,-,+)$.  These are two gauge representatives of the same $BC$ flux sector and are exchanged by the gauge transformation $D_3$.  The direct and indirect paths to $AC$ likewise end in the $D_3$-related representatives $(+,-,-)$ and $(-,+,+)$.

The Berry phase of the effective triangular model in Eq.~(\ref{eq:three_state_Kitaev}) fixes the relative sign of these paths: the two contributions to $BC$ interfere constructively, while those to $AC$ interfere destructively at short times.  In the unprojected Majorana description the paths end in different gauge sectors, so their interference appears only after the two components are recombined coherently.

We first test whether gauge invariance can be restored by applying the gauge projector after the TP evolution, without changing its equations of motion.  We call this procedure post-projection TP (TP$_{\mathrm{pp}}$) and define
\begin{equation}
 \langle O\rangle_{\mathrm{pp}}
 =\frac{\langle P O P\rangle_{\mathrm{TP}}}{\langle P\rangle_{\mathrm{TP}}}.
 \label{eq:post_projection_observable}
\end{equation}
where $\langle\mathord{\cdot}\rangle_{\mathrm{TP}}$ denotes an expectation value evaluated from the raw TP aRDMs. Here $P$ is the gauge projector in Eq.~(\ref{eq:projector}).  When $O$ is gauge invariant, $[O,P]=0$ and $P^2=P$ allow the numerator to be evaluated as $\langle OP\rangle$.

For the $3xyz$ quench, the Zeeman field changes only the three links incident on site 3.  The two gauge representatives generated by the competing spin processes are therefore related by $D_3$, while the other gauge generators act only on spectator links and contribute common factors that cancel in the final expectation value.  It is therefore sufficient here to use $P_3=(1+D_3)/2$.  For the gauge-invariant flux $W_A$,
\begin{equation}
 \langle W_A\rangle_{\mathrm{pp}}
 =\frac{\langle W_AP_3\rangle_{\mathrm{TP}}}{\langle P_3\rangle_{\mathrm{TP}}}
 =\frac{\langle W_A\rangle_{\mathrm{TP}}+\langle W_AD_3\rangle_{\mathrm{TP}}}
 {1+\langle D_3\rangle_{\mathrm{TP}}}.
 \label{eq:post_projection_3xyz}
\end{equation}
In the present geometry $W_A=-u_{43}^{z}u_{36}^{y}$ after inserting the four conserved spectator-link eigenvalues, whose product is $+1$ in the chosen representative, and $\langle W_AD_3\rangle_{\mathrm{TP}}=+\langle b_4^zb_6^yb_3^xc_3\rangle_{\mathrm{TP}}$ is quartic. Therefore, no decoupling is required when evaluating the averages $\langle W_A\rangle_{\mathrm{TP}}$, $\langle W_AD_3\rangle_{\mathrm{TP}}$, and $\langle D_3\rangle_{\mathrm{TP}}$ appearing in Eq.~(\ref{eq:post_projection_3xyz}).

For the $3xyz+5xyz$ quench we instead use $P_{35}=(1+D_3)(1+D_5)/4$.  Expanding $\langle W_AP_{35}\rangle/\langle P_{35}\rangle$ introduces terms such as $\langle D_3D_5\rangle$ and $\langle W_AD_3D_5\rangle$, which reach eight Majorana operators and lie outside the propagated $M^{(1)}(t), M^{(2)}(t)$. We evaluate them by generalizing the decoupling rule in Eq.~(\ref{eq:Cdecouple}) to four particle aRDMs \footnote{In the same normalized-antisymmetrizer convention as Eq.~(\ref{eq:decouple}), setting the connected three- and four-particle aRDMs to zero gives
$M^{(4)}_{abcdefgh}\approx\Upsilon_{abcdefgh}\!\left(35M^{(2)}_{abcd}M^{(2)}_{efgh}-210M^{(1)}_{ab}M^{(1)}_{cd}M^{(1)}_{ef}M^{(1)}_{gh}\right)$.
The coefficients count the distinct $4+4$ partitions and pairings of the eight indices.}.

\begin{figure}[t]
    \centering
    \begin{tabular}{@{}c@{}}
        \begin{overpic}[width=0.86\linewidth]
            {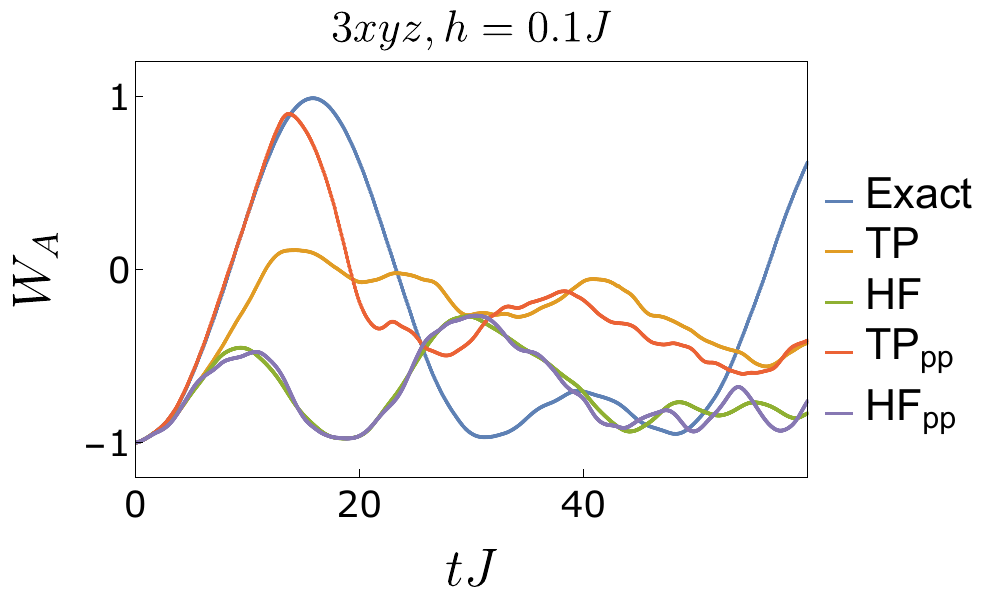}
            \put(0,0){(a)}
        \end{overpic}
        \vspace{0.2cm} \\
        \begin{overpic}[width=0.86\linewidth]
            {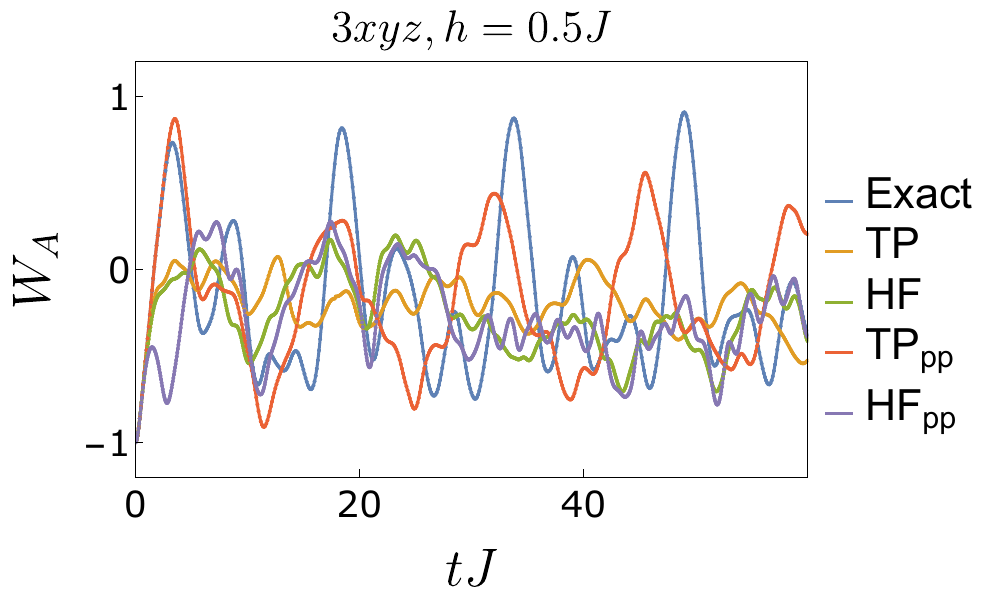}
            \put(0,0){(b)}
        \end{overpic}
    \end{tabular}
    \caption{Post-projection correction after the $3xyz$ evolution.  Exact, TP, HF, TP$_{\mathrm{pp}}$, and HF$_{\mathrm{pp}}$ flux dynamics are compared for (a) $h=0.1J$ and (b) $h=0.5J$.  Both corrections apply $P_3=(1+D_3)/2$ to raw TP and HF results when evaluating the final observable.}
    \label{fig:gauge_methods_3xyz}
\end{figure}

\begin{figure}[t]
    \centering
    \begin{tabular}{@{}c@{}}
        \begin{overpic}[width=0.86\linewidth]
            {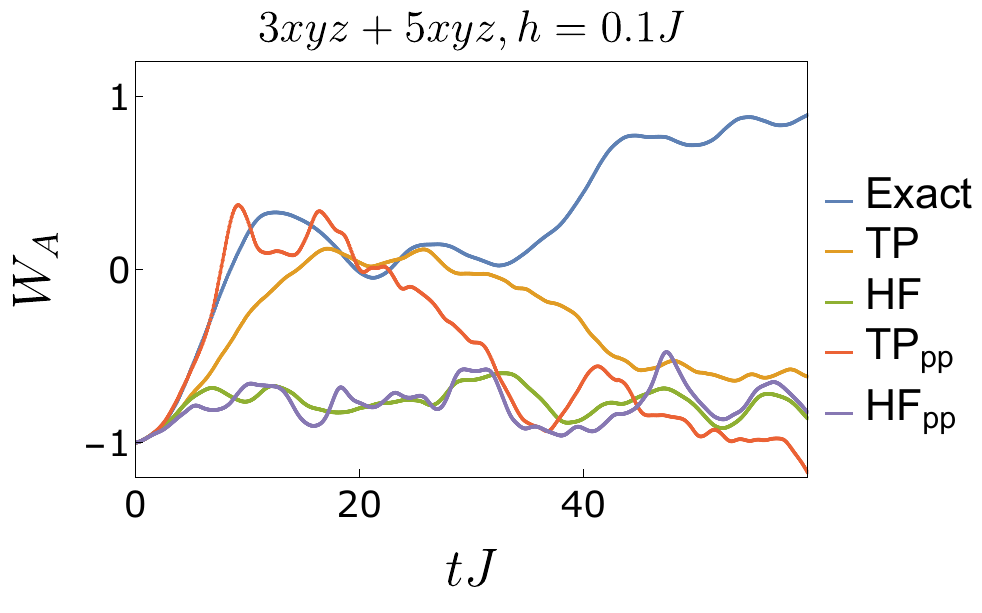}
            \put(0,0){(a)}
        \end{overpic}
        \vspace{0.2cm} \\
        \begin{overpic}[width=0.86\linewidth]
            {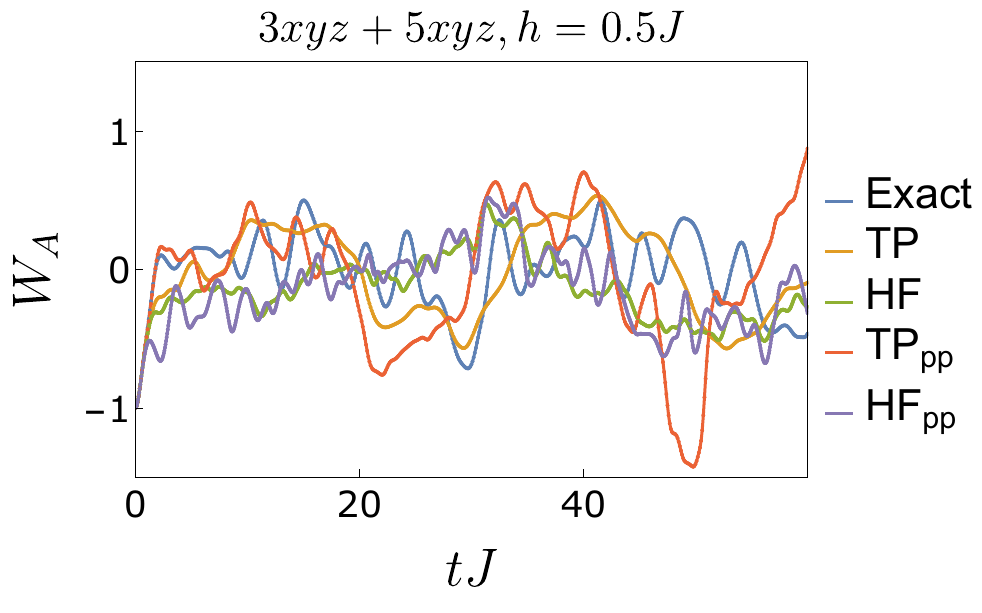}
            \put(0,0){(b)}
        \end{overpic}
    \end{tabular}
    \caption{Post-projection correction after the $3xyz+5xyz$ evolution.  Exact, TP, HF, TP$_{\mathrm{pp}}$, and HF$_{\mathrm{pp}}$ flux dynamics are compared for (a) $h=0.1J$ and (b) $h=0.5J$.  The correction applies the gauge projector $P_{35}=(1+D_3)(1+D_5)/4$ to the raw TP and HF results. The correction procedure eventually leads to unphysical oscillations for both low and high fields, as evidenced by the dip $W_A<-1$ at late times in both curves.}
    \label{fig:gauge_methods_3xyz5xyz}
\end{figure}

We define post-projection HF (HF$_{\mathrm{pp}}$) analogously by applying the same projector to the Gaussian HF state,
\begin{equation}
 \langle O\rangle_{\mathrm{HF,pp}}
 =\frac{\langle P O P\rangle_{\mathrm{HF}}}{\langle P\rangle_{\mathrm{HF}}}.
 \label{eq:hf_post_projection_observable}
\end{equation}
All Majorana averages entering this expression are evaluated from the HF one-particle aRDM by Wick's theorem.

For the single-site quench, Fig.~\ref{fig:gauge_methods_3xyz}, TP$_{\mathrm{pp}}$ recovers the initial rise and first maximum far more accurately than TP.  At low field it still misses the later revival, while at high field it improves several peaks without reproducing their full amplitudes.  By contrast, HF$_{\mathrm{pp}}$ makes no improvement over HF; the low field HF$_{\mathrm{pp}}$ curve tracks the raw HF curve. This suggests that raw HF evolution retains only one of the two $D_3$-related gauge representatives, whereas TP retains information from both.  Because post-projection can only recombine components already present in the approximate state, it has little to act on in HF.

The multisite result in Fig.~\ref{fig:gauge_methods_3xyz5xyz} is less controlled.  TP$_{\mathrm{pp}}$ improves the early evolution, following the exact low-field curve for approximately $t<7J^{-1}$ and the high-field curve for approximately $t<5J^{-1}$, but later develops large, irregular oscillations. In fact, the valley near $t=48J^{-1}$ dips below $\braket{W_A}<-1$, which is unphysical.  HF$_{\mathrm{pp}}$ again improves very little on HF.  As the state evolves, it develops connected three- and four-particle correlations.  These are discarded by the TP closure, so the reconstruction of higher-order correlators becomes less accurate.  The single-site post-projection requires only the retained two-particle aRDM, whereas $P_{35}$ requires four-particle aRDMs. The resulting four-particle observables therefore carry substantially larger errors, leading to less controlled dynamics.

This limitation grows with the number of sites on which the field acts.  If there are $m$ relevant gauge transformations, then the projector $P_{m}=2^{-m}\prod_{j=1}^m(1+D_j)$ contains $2^m$ products and, before geometry-dependent cancellations, reaches Majorana rank of order $4m$. Therefore, we expect the time evolution to be less and less accurate as the field is applied to more sites.

A second strategy is to impose gauge invariance during the TP evolution.  For the $3xyz$ quench we enforce the $D_3$ identity
\begin{equation}
 \langle O\rangle=\langle D_3O\rangle
 \label{eq:d3_dynamic_constraint}
\end{equation}
whenever both sides belong to the retained one- or two-particle aRDMs, together with the corresponding identities in the reconstructed three-particle aRDM.  Thus $D_3$-odd moments are set to zero, moments paired by multiplication with $D_3$ are identified with the appropriate sign, and reconstructed six-operator expectation values containing all four Majoranas of $D_3$ are replaced by their signed bilinear partners. If $D_3O$ is a three-particle operator while $O$ belongs to the one-particle aRDM, then during reconstruction we replace the value of $\langle D_3O\rangle$ by the retained moment $\langle O\rangle$, with the appropriate sign.  We impose these identities on the initial aRDM, enforce them at every time step, and reimpose them after each purification.  In addition, purification preserves $\langle D_3\rangle$ at its initial value, and we treat it as a conserved quantity alongside the energy. We call the above procedure gauge-invariant TP (TP$_{\mathrm{gi}}$).

\begin{figure}[t]
    \centering
    \begin{tabular}{@{}c@{}}
        \begin{overpic}[width=0.86\linewidth]
            {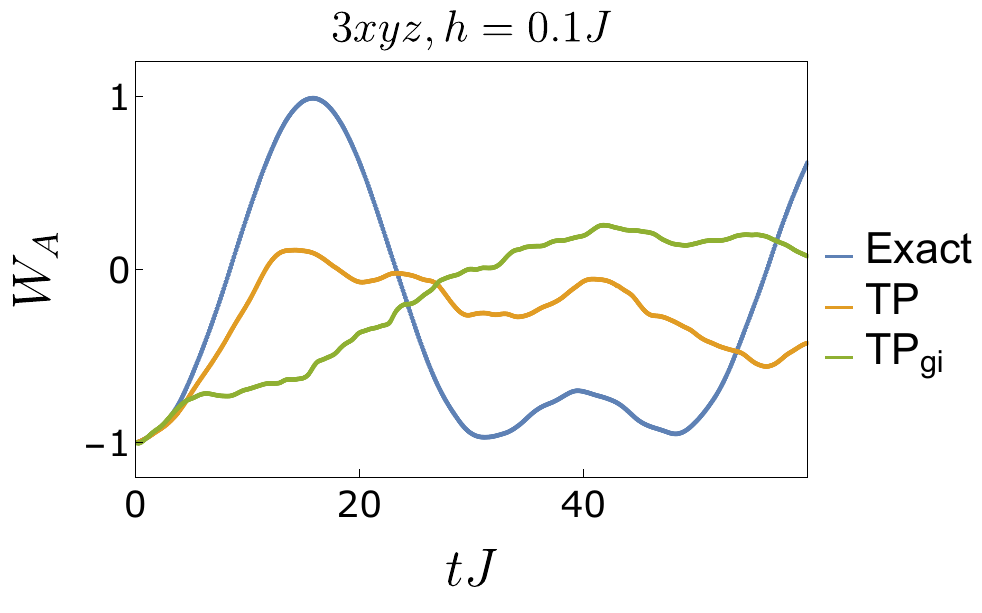}
            \put(0,0){(a)}
        \end{overpic}
        \vspace{0.2cm} \\
        \begin{overpic}[width=0.86\linewidth]
            {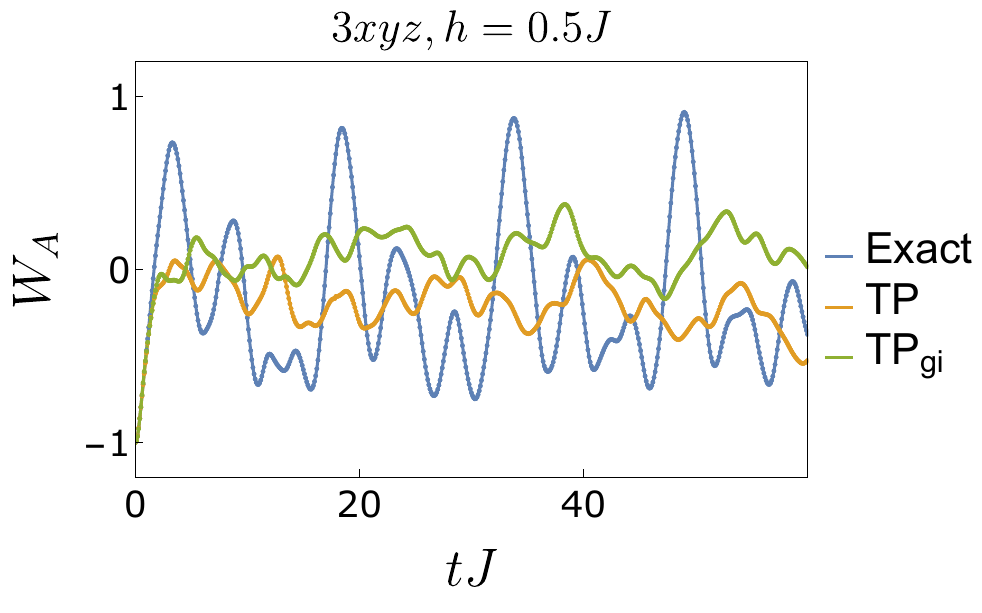}
            \put(0,0){(b)}
        \end{overpic}
    \end{tabular}
    \caption{Gauge constraints imposed during the $3xyz$ TP evolution.  The exact, TP, and gauge-invariant TP$_{\mathrm{gi}}$ flux dynamics are compared for (a) $h=0.1J$ and (b) $h=0.5J$.  TP$_{\mathrm{gi}}$ enforces the gauge constraint Eq.~(\ref{eq:d3_dynamic_constraint}) at every step in the time evolution.}
    \label{fig:gauge_constrained_3xyz}
\end{figure}

Figure~\ref{fig:gauge_constrained_3xyz} shows no systematic improvement.  At low field TP$_{\mathrm{gi}}$ misses the peak near $t=16J^{-1}$. Indeed, its value there is even lower than that predicted by raw TP. At high field it oscillates around a higher mean value than TP and does not reproduce the four peaks of the exact dynamics.

The failure of TP$_{\mathrm{gi}}$ can be understood from the solvable Kitaev limit.  In a fixed gauge sector, the ground state is Gaussian and is completely specified by its correlation matrix $\langle c_i c_j\rangle$.  These bare correlators are not physical gauge-invariant observables: for example, $c_3c_4$ anticommutes with $D_3$ and its expectation must vanish in a $D_3$-invariant description.  Under gauge invariance, the corresponding physical information is transferred to a higher order correlator.  For the nearest-neighbor bond $c_3 c_4$, the natural gauge invariant alternative is $b_3^zb_4^zc_3c_4$, which is the bond-energy operator and is already quartic.  More generally,
\begin{equation}
 \langle i c_i c_j\rangle
 \rightarrow
 \left\langle i c_i
 \left(\prod_{\langle k l\rangle\in\Gamma_{ij}}u_{kl}\right)c_j\right\rangle .
 \label{eq:gauge_string_completion}
\end{equation}
Here $\Gamma_{ij}$ is a path from site $i$ to site $j$. A path $\Gamma_{ij}$ containing $L$ distinct links generally produces $2L+2$ Majorana operators, or an element of the $(L+1)$-particle aRDM.  Thus a low-rank gauge-invariant Majorana description contains less of the information that appears in the correlation matrix of a fixed gauge; that information reappears in higher-order correlators.  As a magnetic field acts on more sites and makes more links dynamical, the required strings and projector moments extend beyond any fixed low-order hierarchy. Consequently, forcing a fixed set of low-order aRDMs to satisfy every gauge identity visible at that order necessarily discards information that would otherwise be present in a fixed-gauge description.

\section{Conclusions}

We have introduced a {time-dependent TP method} for fermionic systems with quadratic and quartic interactions. The method utilizes a hierarchy of equations for the one- and two-particle aRDMs. Truncating this hierarchy at the two-particle level gives a natural extension of time-dependent HF: the HF approximation is recovered when the two-particle aRDM is factorized, while the present method keeps the connected two-particle correlations explicitly.

By giving a schematic process of reconstructing the full quantum state from the two-particle aRDM, we show that the equation hierarchy, even if truncated, still conserves average values for conserved operators expressible as sums of products of no more than four Majorana operators. At the same time, this reconstructed state is not guaranteed to be positive, which leads to instabilities that can be controlled by certain projection processes. These projections can be made so that the conserved quantities remain unchanged. The price is that the projection is not a unitary operation: it removes information from the reduced description and makes the long-time dynamics effectively irreversible.

The numerical examples show both the usefulness and the limitations of this approach. For the one-dimensional Hubbard model, TP captures the dynamics substantially better than HF in the regime of strong interactions. At longer times, however, the repeated positivity projection drives the solution toward an equilibrium-like state, with a rate that depends on how often the projection is applied. The perturbative correction to the closure, while costing significantly more compute [$O(N^8)$], does not provide a substantial improvement.

The Kitaev-model calculations provide a second test in which two-particle correlations and gauge structure are both important. For Zeeman fields where restoring gauge invariance is unnecessary, TP dynamics is near exact over a reasonably long time. For the multicomponent low-field quenches, after restoring gauge invariance by applying the gauge projector, TP$_\mathrm{pp}$ accurately reproduces the initial time evolution. However, for a Zeeman field applied to more than one site, TP$_\mathrm{pp}$ develops large, uncontrolled oscillations at late times.

The Kitaev tests also reveal a limitation that is specific to gauge-redundant Majorana descriptions. When the gauge fields strongly fluctuate, restoring gauge invariance requires costly long Majorana strings. A direct spin representation, in which gauge invariance is automatic, may then be more efficient.

Our results suggest that TP dynamics provides a useful intermediate description between mean-field theory and exact many-body evolution. The method propagates only one- and two-particle objects, while retaining the two-particle correlations that are important in strong-coupling Hubbard transport and field-induced flux motion in the Kitaev model. Its main limitation is computational cost: the present algorithm scales as $O(N^6)$. In addition, the projections used to stabilize the evolution introduce irreversibility into the dynamics. The resulting projection-induced thermalization is therefore most benign when it occurs on timescales longer than the correlated dynamics of interest, or in systems whose exact dynamics already thermalizes. For systems with persistent coherent oscillations, improving the projection procedure will be an important direction for future work.

\begin{acknowledgments}
We acknowledge support from the Leverhulme Trust under the grant agreement RPG-2023-253.

OpenAI Codex (GPT-5.6 Sol and GPT-6 Astra) was used under the authors' direction to assist with manuscript revision, derivation checks, and the development and checking of numerical-analysis code. The authors reviewed and revised the resulting material and verified the adopted suggestions. The authors take full responsibility for the scientific content and conclusions of this work.
\end{acknowledgments}

\appendix
\section{High-field test of the flux-sector effective model}
\label{app:effective_high_field}

\begin{figure}[t]
    \centering
    \begin{tabular}{@{}c@{}}
        \begin{overpic}[width=0.95\linewidth]
            {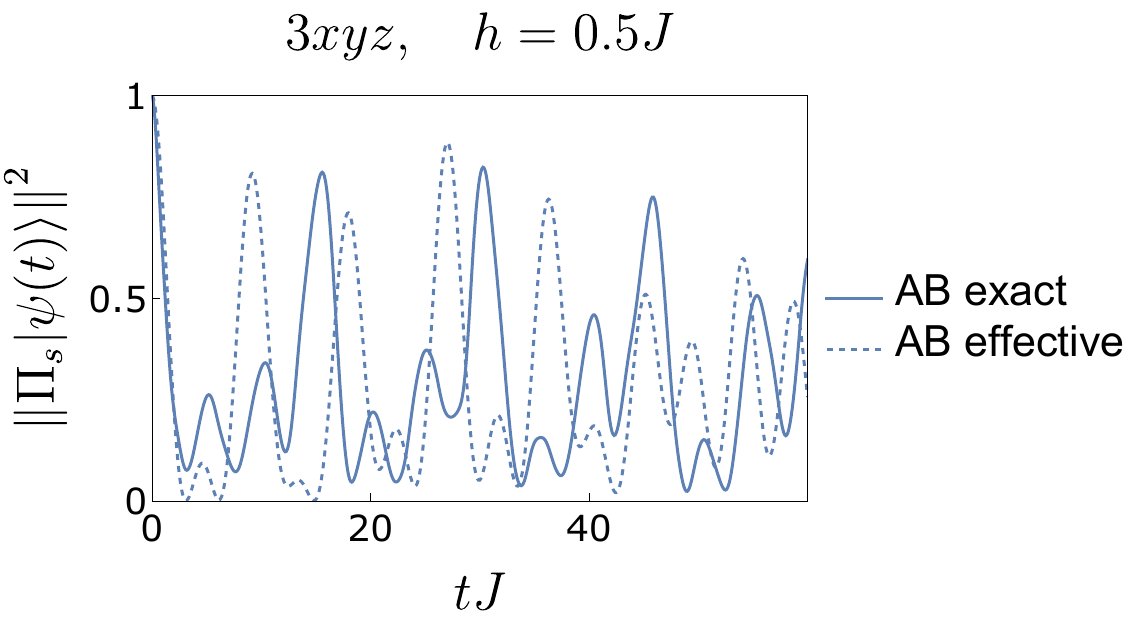}
            \put(0,0){(a)}
        \end{overpic}
        \vspace{0.2cm} \\
        \begin{overpic}[width=0.95\linewidth]
            {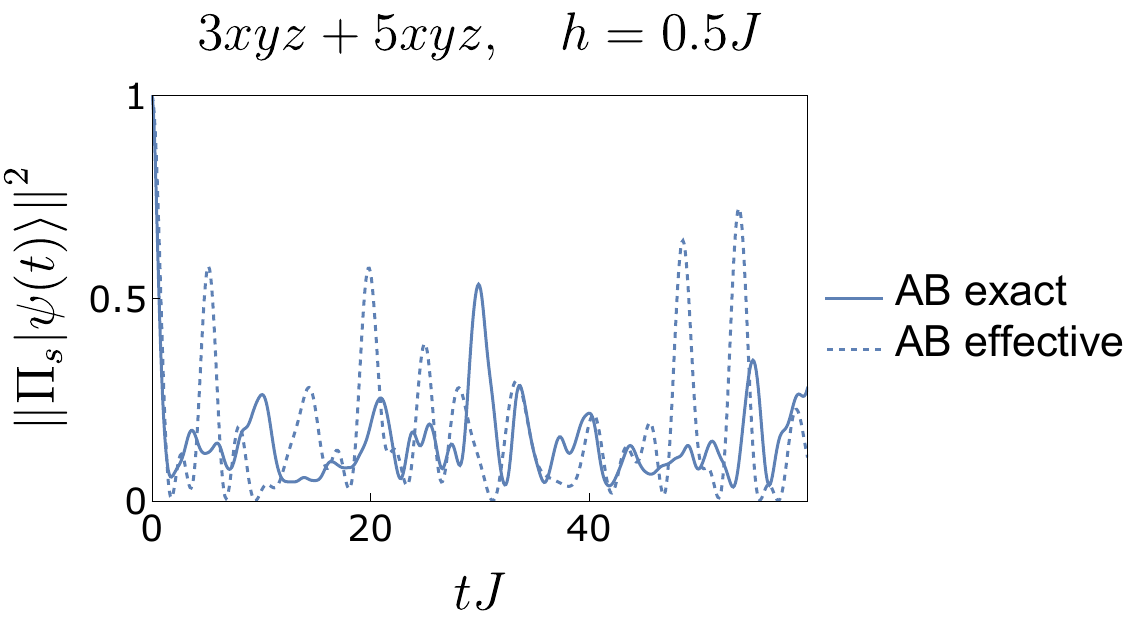}
            \put(0,0){(b)}
        \end{overpic}
    \end{tabular}
    \caption{High-field test of the branch-resolved effective dynamics at $h=0.5J$. The solid blue curve is the exact $AB$-sector probability and the dashed blue curve is the corresponding effective-model result. All flux sectors connected to $AB$ are retained. (a) $3xyz$ quench. (b) $3xyz+5xyz$ quench. }
    \label{fig:effective_high_field_appendix}
\end{figure}

We test the flux-sector effective model at $h=0.5J$ in Fig.~\ref{fig:effective_high_field_appendix}. Within each flux sector $s$ we retain the ground state $\ket{s}$ in every reachable flux sector and define
\begin{equation}
(H_{\mathrm{eff}})_{ss'}
=E_{s}\delta_{ss'}
+\langle s|H_h|s'\rangle.
\end{equation}
For the $3xyz$ quench the reachable flux sectors are $\{FF,AB,AC,BC\}$, whereas the $3xyz+5xyz$ field connects all eight flux sectors $\{FF,AB,\ldots,CD,ABCD\}$. To keep the high-field comparison legible, Fig.~\ref{fig:effective_high_field_appendix} shows only the $AB$-sector probability.

For the $3xyz$ quench, the effective model follows the initial decrease of the exact $AB$-sector probability, but reaches its first minimum and subsequent small maximum too early. At later times, its amplitudes disagree with the exact result, and its oscillations are generally too rapid. The disagreement is more pronounced for the $3xyz+5xyz$ quench, where the effective model predicts several large peaks that are absent from the exact dynamics.

\bibliographystyle{apsrev4-2}
\bibliography{biblio}

\end{document}